\documentclass{aa}
\usepackage[utf8]{inputenc} 
\usepackage[colorlinks=true, citecolor=blue, linkcolor=blue]{hyperref}
\usepackage{cases}
\usepackage[english]{babel}
\usepackage{multicol}
\usepackage{amsmath}
\usepackage{amsfonts}
\usepackage{dsfont}
\usepackage{amsxtra}
\usepackage{amssymb}
\usepackage{upgreek} 
\usepackage{changes}
\usepackage{multirow}
\usepackage{url}
\usepackage{float}

\usepackage{graphicx,epsfig}
\usepackage{dcolumn}

\usepackage{xspace} 

\usepackage{color}
\usepackage{xcolor}

\usepackage[switch]{lineno}
\usepackage{lipsum}

\begin{document}

\title{$\gamma$-Cygni supernova remnant in $\gamma$-rays: signatures of trapped and escaped Cosmic Rays }

 \author{Yuan Li\thanks{yuanlss17@sjtu.edu.cn}
	\inst{1,2}
	\and
	Gwenael Giacinti\inst{1,2}\thanks{gwenael.giacinti@sjtu.edu.cn}
	\and
	Siming Liu\thanks{liusm@swjtu.edu.cn}
	\inst{3}
	\and
	Yi Xing
	\inst{4}
}

\institute{Tsung-Dao Lee Institute, Shanghai Jiao Tong University, Shanghai 201210, PRC
	\and
	{School of Physics and Astronomy, Shanghai Jiao Tong University, Shanghai 200240, PRC}
	\and
	School of Physical Science and Technology, Southwest Jiaotong University, Chengdu 610031, PRC
	\and
	Key Laboratory for Research in Galaxies and Cosmology, Shanghai Astronomical Observatory, Chinese Academy of Sciences, Shanghai 200030, PRC \\
}

\abstract{We reanalyze 15 years of data recorded by the Fermi Large Area Telescope in a region around supernova remnant (SNR) $\gamma$-Cygni from 100 MeV to 1 TeV, and find that the spectra of two extended sources associated with the southeast radio SNR arc and the TeV VERITAS source can be described well by single power-laws with photon indices of $2.149\pm0.005$ and $2.01\pm0.06$, respectively. Combining with high resolution observations of surrounding gas, we model the emission in the hadronic scenario, where the $\gamma$-ray emission could be interpreted as escaped cosmic ray (CR) illuminating a nearby Molecular Cloud (MC) plus an ongoing shock-cloud interaction component. In this scenario, the difference between the two GeV spectral indices is due to the different ratios of MC mass between the escaped component and the trapped component in the two regions. We further analyze, in a potential pulsar halo region, the relationship between energy density $\varepsilon_{\rm{e}}$, spin-down power $\dot{E}$, and the $\gamma$-ray luminosity $L_{\gamma}$ of PSR J2021+4026. Our results indicate that the existence of a pulsar halo is unlikely. On the other hand, considering the uncertainty on the SNR distance, the derived energy density $\varepsilon_{\rm{e}}$ might be overestimated, thus the scenario of a SNR and a pulsar halo overlapping in the direction of the line of sight (LOS) cannot be ruled out.}


\maketitle

\section{Introduction} \label{sec:intro}
Supernova remnants (SNRs) are widely recognized as the leading candidates for accelerating Galactic cosmic rays (CRs) up to energies below the so-called "knee" in the CR spectrum, around the PeV scale \citep{Hillas2005}. In these extreme environments, high-energy CRs give rise to $\gamma$-ray emission through different radiative processes. In the leptonic scenario, $\gamma$-rays are produced via inverse Compton scattering or bremsstrahlung of relativistic electrons. Conversely, the hadronic scenario involves inelastic collisions between accelerated protons and ambient matter, resulting in the production of neutral pions, which decay into $\gamma$-rays. A particularly distinctive signature of hadronic CR acceleration is the detection of $\gamma$-ray emission from dense molecular clouds (MCs) illuminated by CRs that have escaped nearby SNRs. Observational support for this scenario has been provided by the \emph{Fermi} Large Area Telescope (\emph{Fermi}-LAT), which has reported $\gamma$-ray emission consistent with CR–MC interactions in several well-known SNRs, including IC 443 \citep{Ackermann2013, Abdo2010observation}, W44 \citep{uchiyama2012fermi, peron2020gamma}, W28 \citep{aharonian2008discovery, li2010gamma, hanabata2014detailed}, W51C \citep{abdo2009fermi}, G150.3+4.5 \citep{2024A&A...689A.257L}, and G15.4+0.1 \citep{2023ApJ...945...21L}. These remnants exhibit prominent GeV $\gamma$-ray emission, generally attributed to hadronic interactions between shock-accelerated protons and nearby dense gas. In particular, the $\gamma$-ray spectra of W44 and W51C reveal a characteristic spectral bump associated with neutral pion decay—providing some of the most compelling evidence to date for the acceleration of relativistic protons in SNRs \citep{2011ApJ...742L..30G, 2016ApJ...816..100J}.

The SNR $\gamma$-Cygni (G78.2+2.1) is a puzzling $\gamma$-ray source located in the complex Cygnus region. The multi-wavelength analysis towards this well-studied region have never ceased, and the origins of its $\gamma$-ray emission is still not well understood. \citet{2008A&A...490..197L} revealed the radio flux spectral index $-\alpha_v$ varies between $\sim$ -0.8 and $\lesssim$ -0.4 across the whole SNR region. The softest index is found in the bright south-eastern part, and the north-west part is harder ($\sim$ 0.55) \citep{2008A&A...490..197L,2023A&A...670A...8M}. Near the center, there is an energetic GeV-bright pulsar PSR J2021+4026, for which it has been proven that the pulse profile varies~\citep{2023arXiv230703661W,2023A&A...676A..91R}. Also, the unknown distances make the $\gamma$-ray emission in this region more puzzling. In the very-high-energy (VHE) energy band, VERITAS observations report a $\sim 0.23\degr$ extended source in the northwest part \citep{2013ApJ...770...93A}, which is characterized using an ellipse template ($0.29\degr\times 0.19\degr$) with a power-law index of 2.79 $\pm 0.39_{\rm stat} \pm 0.20_{\rm sys}$ \citep{2018ApJ...861..134A}. The VERITAS source location is roughly consistent with part of the MAGIC source \citep{2023A&A...670A...8M}, while both of them are spatially inconsistent with 2HWC J2020+403 measured by HAWC \citep{2017ApJ...843...40A}. Furthermore, \citet{2023A&A...670A...8M} found that there is a bright arc-like TeV source located on the west side outside the SNR, which is not detected in the GeV band, and makes the origins of the $\gamma$-ray emission in this region even more puzzling. 

Pulsar halos are a new type of $\gamma$-ray source identified in recent years. As Pulsar Wind Nebulae (PWN) age, high-energy electrons and positrons may escape from the nebula and diffuse in the surrounding interstellar medium (ISM), after $\sim 100$\,kyr \citep{2020A&A...636A.113G}. The electrons and positrons diffusing out of the PWN produce $\gamma$-ray emission through Inverse Compton scattering of ambient photon fields \citep{2022NatAs...6..199L}. The recent detection of extended $\gamma$-ray emission around Geminga and Monogem\citep{2017ApJ...843...40A,2017Sci...358..911A,2020PhRvD.101j3035D}, and the slow diffusion region around them, has sparked a widespread discussion about TeV halos, while the nature of the emission is still uncertain. It is also noted that pulsar halos may exhibit an asymmetric morphology when the coherence length is large enough \citep{2018MNRAS.479.4526L}, indicating the potential anisotropic diffusion inside the pulsar halo. However, limited by the resolution and systematic uncertainties, an asymmetric morphology of these pulsar halos has not been confirmed by experiments yet \citep{2017Sci...358..911A, 2023A&A...673A.148H}.
%\citet{2020PhRvD.101j3035D} argue that extended GeV emission can be measured with \emph{Fermi} satellite around, while this is not confirmed by \citet{2019ApJ...878..104X}

In this work, we conduct a comprehensive analysis of the GeV $\gamma$-ray emission towards the $\gamma$-Cygni area, utilizing 15 years of data from the \emph{Fermi}-LAT instrument. The photon events are selected in an energy range from 100 MeV to 1 TeV, and the results are presented in Sect. \ref{sec:2}. In Sect. \ref{sec:3}, we presents the gas observation results of $^{12}$CO(J = 1-0) obtained from the Milky Way Imaging Scroll Painting (MWISP). Sect. \ref{sec:4} is dedicated to discussing potential origins of the $\gamma$-ray emission, comparing multi-wavelength observations with theoretical expectations. Lastly, Sect. \ref{sec:5} provides our conclusions.

\section{\emph{Fermi}-LAT Data Reduction}\label{sec:2}

In the subsequent analysis, the standard LAT analysis software \emph{Fermitools} v2.2.0 is adopted, together with the \emph{Fermipy} v1.1.6 \citep{2017ICRC...35..824W} to quantitatively calculate the extension and position of extended sources, and the most recent \emph{Fermi}-LAT \citep{Atwood2009} Pass 8 data are collected from August 4, 2008 (Mission Elapsed Time 239557418) to August 4, 2023 (Mission Elapsed Time 712800005) to study the GeV emission around $\gamma$-Cygni. 
We select the data with "Source" event class ``P8R3$\_$SOURCE'' (evclass=128) and event type FRONT + BACK (evtype=3), with the standard data quality selection criteria $\tt (DATA\_QUAL > 0)  \&\& (LAT\_CONFIG == 1)$, and exclude the zenith angle over 90$\degr$ to avoid the earth limb contamination. To derive a better point-spread function and reduce contamination from the pulsar, photon energies are limited to between 20 GeV and 1 TeV for further morphological analysis. Additionally, events with energies between 100 MeV and 1 TeV are selected for a more detailed spectral analysis. All photon events within a $20\degr\times20\degr$ region of interest (ROI) centered at the position of $\gamma$-Cygni are modeled using the Fermi-LAT 4FGL catalog data release 4 (4FGL-DR4;\citep{2020ApJS..247...33A,2023arXiv230712546B}) in a binned maximum likelihood analysis \citep{mattox1996likelihood}. The instrument response functions (IRF) ``P8R3\_SOURCE\_V3''\footnote{\url{http://fermi.gsfc.nasa.gov/ssc/data/access/lat/BackgroundModels.html}}, along with the Galactic/isotropic diffuse background models (IEM, $\tt gll\_iem\_v07.fits$)/($\tt iso\_P8R3\_SOURCE\_V3\_v1.txt$ ) are adopted. All sources listed in the 4FGL-DR4 catalog are included in the background model, all sources within $7\degr$ from the center of ROI and two diffuse backgrounds above are set free, which is generated by the script make4FGLxml.py\footnote{\url{ http://fermi.gsfc.nasa.gov/ssc/data/analysis/user}}. The likelihood test statistic (TS) is adopted to calculated the significance of the $\gamma$-ray sources, which is defined as TS$= 2 (\ln\mathcal{L}_{1}-\ln\mathcal{L}_{0})$, where $\mathcal{L}_{1}$ and $\mathcal{L}_{0}$ are maximum likelihood values for the model with target source and without target source. Furthermore, the $\rm TS_{ext}$ is defined as $\rm {TS_{ext}} = 2(\ln\mathcal{L}_{\rm ext} - \ln\mathcal{L}_{\rm ps})$, where $\mathcal{L}_{\rm ext}$ and $\mathcal{L}_{\rm ps}$ represent the maximum likelihood values for the extended and point-like templates, respectively. This calculation considers only one additional free parameter introduced by the extended template, and the significance is approximately given by $\sqrt{\rm TS_{ext}}$ in units of $\sigma$.

\subsection{Timing Analysis}

\begin{figure}
    \centering
    \includegraphics[trim={0 0.cm 0 0}, clip,width=0.4\textwidth]{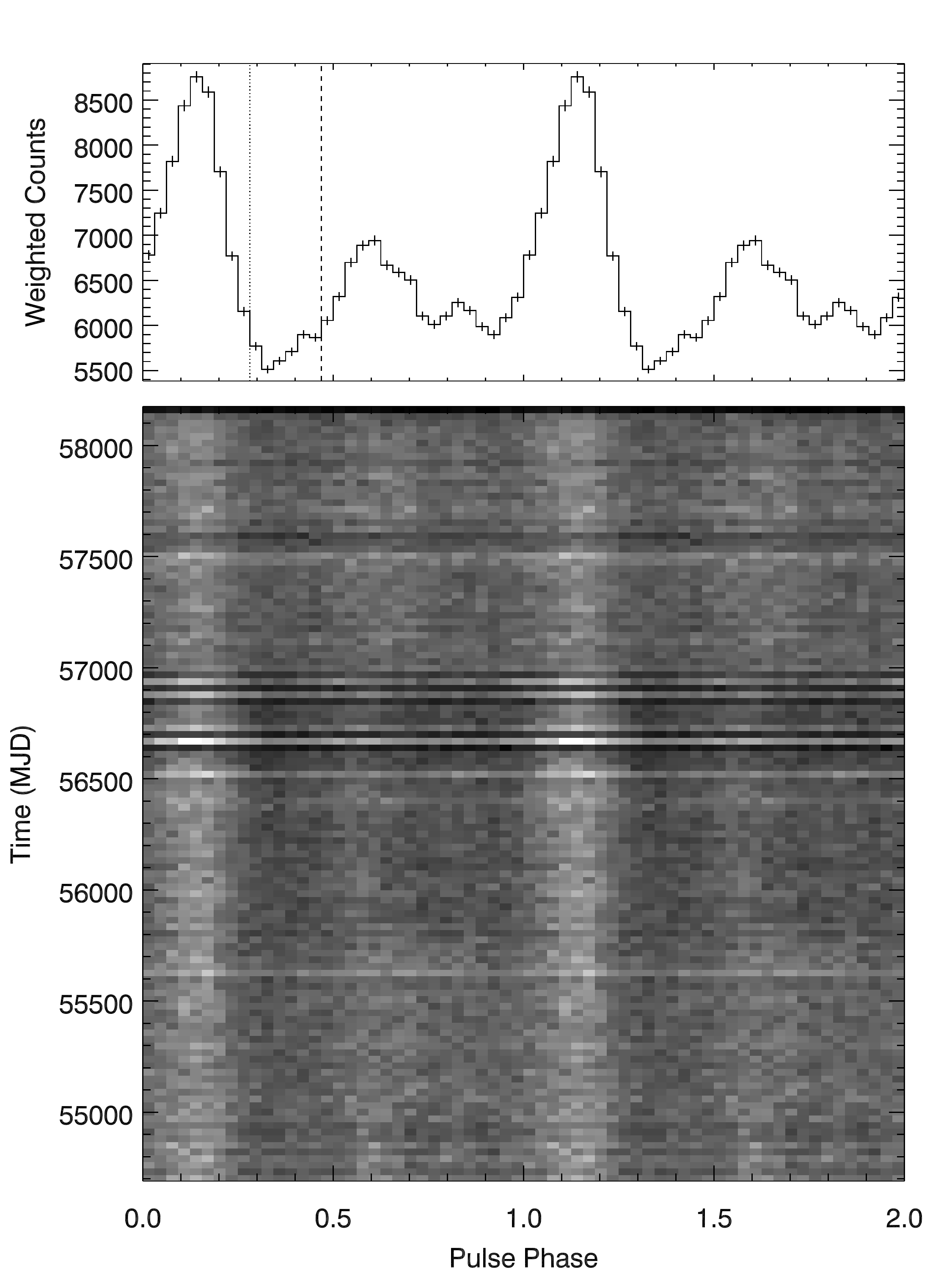}
    \caption{Folded pulse profile and two-dimensional phaseogram in 32 phase bins derived for PSR J2021+4026. The greyscale represents the weights of photons in each bin white = higher weight, black = lower weight, and the dotted and dashed lines mark the minimum and maximum phases of the off-pulse phase interval, respectively.}
   \label{fig:0}
 \end{figure}

We performed a timing analysis to the 0.1-1000 GeV LAT data of the PSR J2021+4026 to separate its $\gamma$-ray emission from other sources in the region. We first included the LAT photons within an aperture radius of 3$\degr$, and weighted them by their probability of originating from the pulsar (using gtsrcprob in the Fermitools). The LAT photons with low weights ($<$0.001) were excluded in the analysis.
We assigned pulse phases to the LAT photons using the Fermi plugin of TEMPO2 \citep{2006MNRAS.369..655H,2006MNRAS.372.1549E}, according to the known ephemeris given in the LAT third pulsar catalog \citep{smi+23}. This ephemeris covers the time period of MJD 54689--58175, so we only included the LAT data during this time period. The weighted pulse profile and the two-dimensional phaseogram are plotted in Figure \ref{fig:0}. We defined the phase 0.28--0.47 as the off-pulse phase range, and the remaining phase as the on-pulse phase range.

\begin{figure*}
    \centering
    \includegraphics[trim={0 0.cm 0 0}, clip,width=0.321\textwidth]{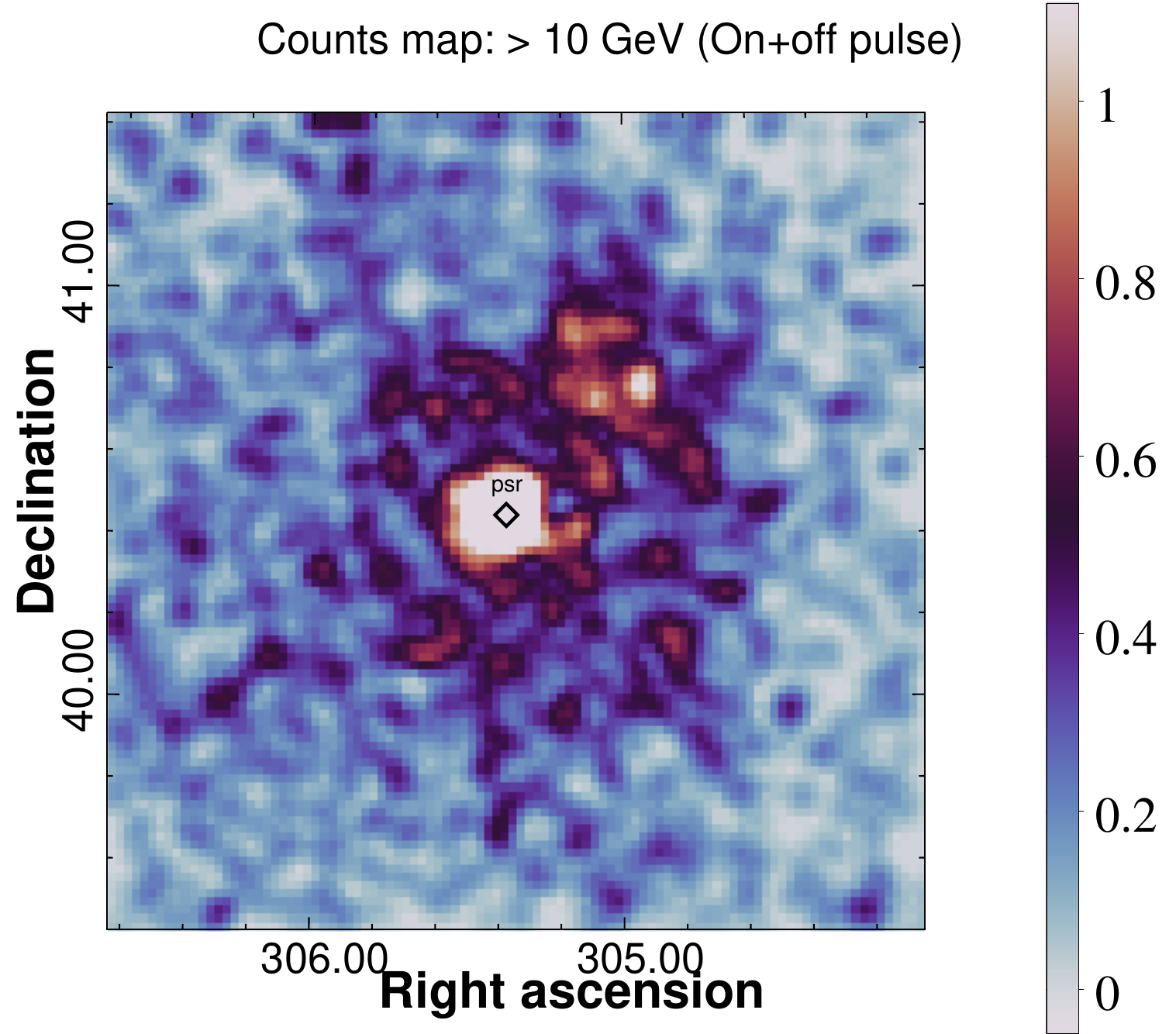}   
    \includegraphics[trim={0 0.cm 0 0}, clip,width=0.319\textwidth]{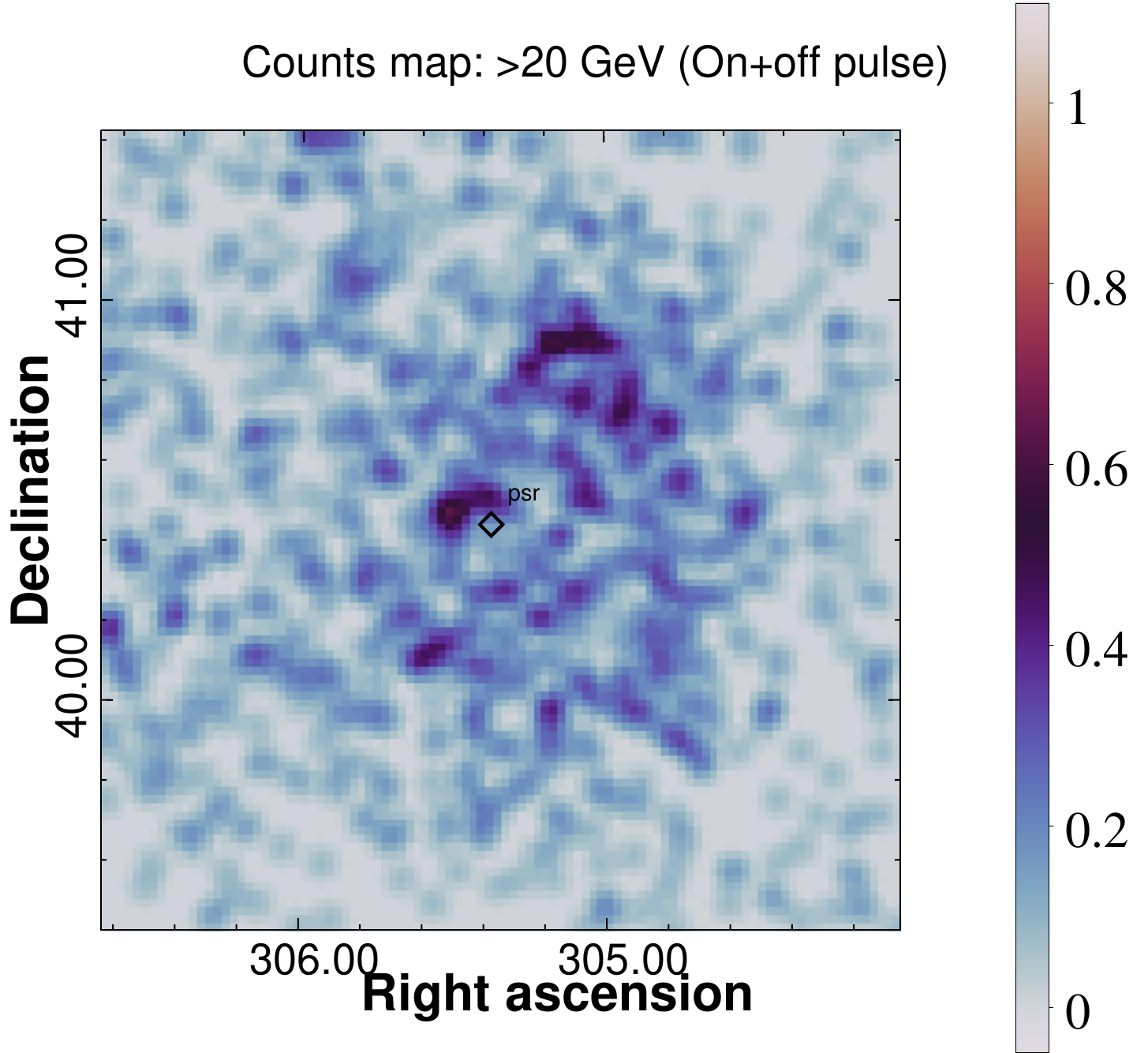} \\
    \includegraphics[trim={0 0.cm 0 0}, clip,width=0.32\textwidth]{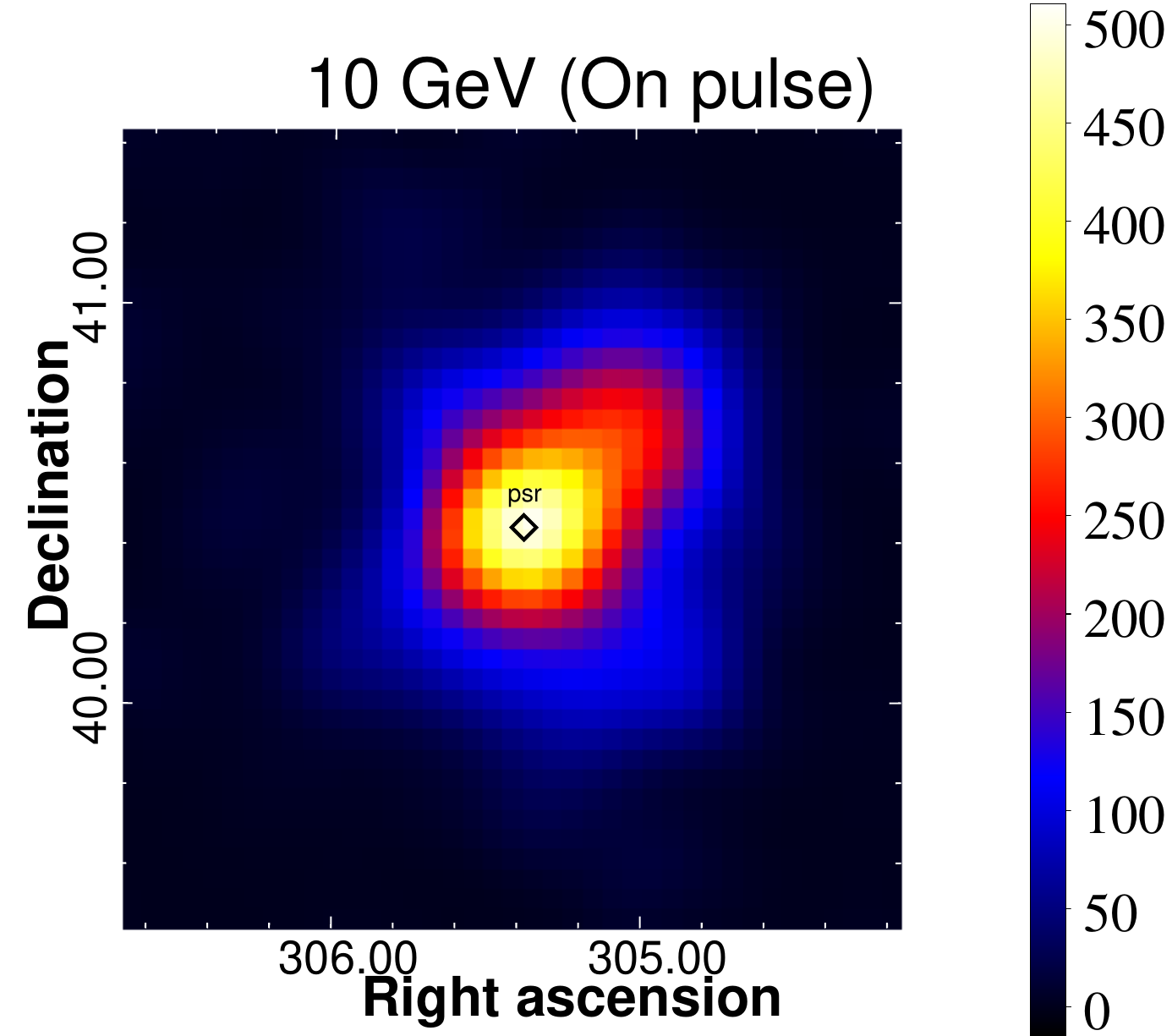}   
    \includegraphics[trim={0 0.cm 0 0}, clip,width=0.327\textwidth]{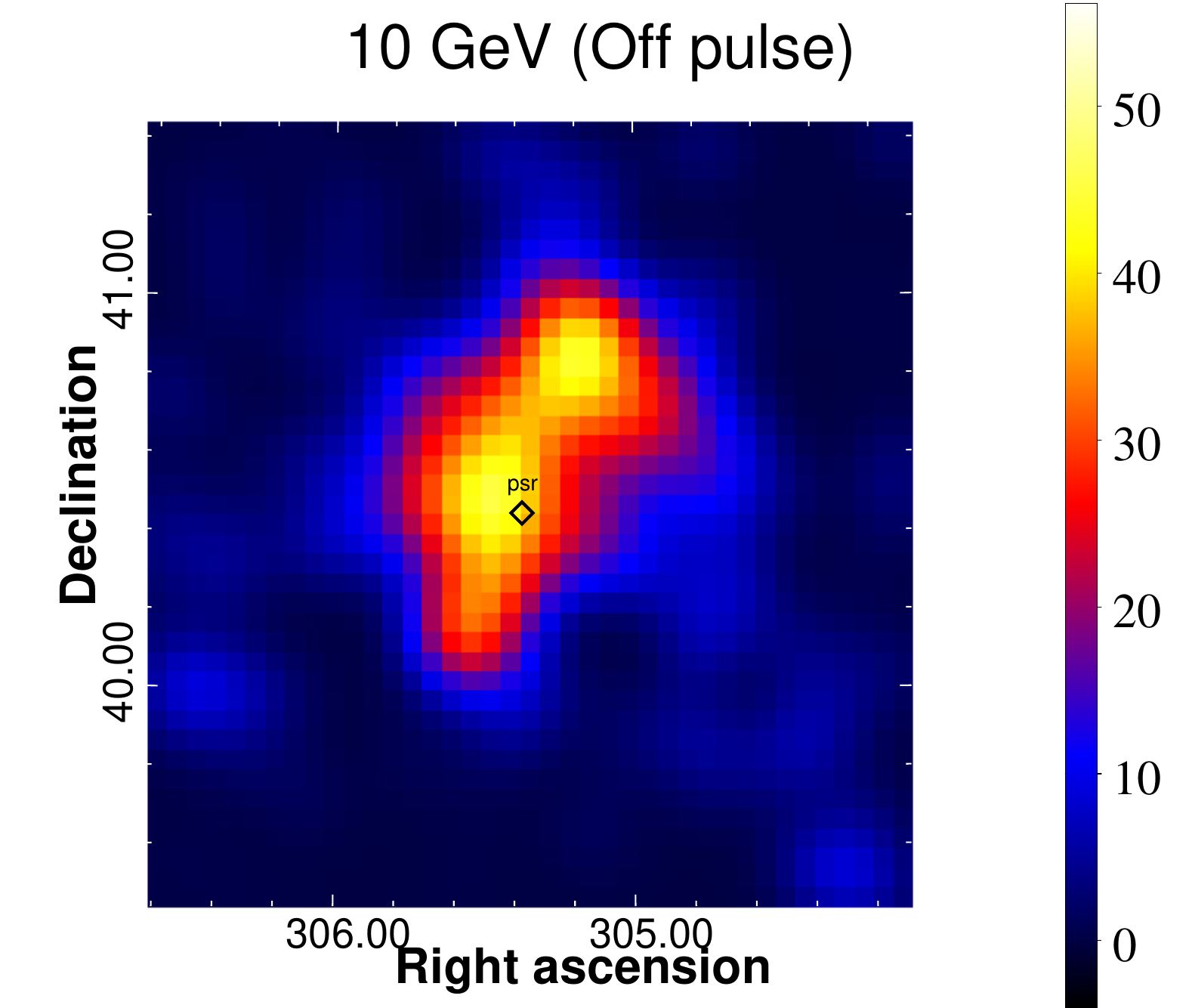} 
    \caption{Top row: \emph{Fermi}-LAT counts map above 10 GeV and 20 GeV within full-phase range, respectively. Bottom row: \emph{Fermi}-LAT TS map above 10 GeV of LAT data during the On-pulse and Off-pulse phases, respectively. The black diamond shows the position of the gamma-ray pulsar PSR J2021+4026 \citep{2009Sci...325..840A}.}
   \label{fig:9}
 \end{figure*}

Then we assigned pulse phases to all of the LAT events in the ROI.
With LAT events during different phase ranges, we generated counts maps in different energy ranges to verify if we can avoid the contamination from PSR J2021$+$4026. The resulting counts map are depicted in the top row of Figure \ref{fig:9}.
From the top left panel of Figure \ref{fig:9}, it can be seen that most $>$10 GeV photon events are concentrate around PSR J2021$+$4026 within the full-phase (including both on-pulse and off-pulse) range. However, almost no $\gamma$-ray excess is detected when $>$20 GeV photon events are selected, as shown in the top right panel of Figure \ref{fig:9}. Additionally, we calculated the TS maps in the vicinty of PSR J2021$+$4026 for both the on-pulse (bottom left panel of Figure \ref{fig:9}) and off-pulse phase using $>$10 GeV photon events. The significant excess around the pulsar is still present even when we only included events during the off-pulse phase range (shown as the bottom right panel of Figure \ref{fig:9}). These results indicate that it is not enough to separate the pulsar's contamination in $>$10 GeV energy band, in which energy band, the TS value of the pulsar is still more than 500, while when we adopted $>$20 GeV data events, the TS value is only $\sim$4, in contrast to $\sim$249,000 measured with $>$100 MeV data events. Therefore, we adopted $>$20 GeV full-phase range data events to conduct a quantitative analysis of the spatial components within the region. 

\subsection{Morphological analysis}\label{sec:2.2}

In the beginning, we attempt to verify the $\gamma$-ray emission position offset with energy, and also the energy-dependent morphology. Combined with the above results, we only select photon events above 20 GeV to generate TS maps, and all TS maps are generated by only considering background fitting but not including the extended source $\gamma$-Cygni. The $\gamma$-ray contribution from PSR J2021+4026 has been removed, which is modeled by a point-like source, whose best-fit location in the full-phase range given by \emph{gtfindsrc} is marked as R.A. = $305.384^{\circ}\!\pm0.009^{\circ}\!$, Dec. = $40.441^{\circ}\!\pm0.012^{\circ}\!$, which is consistent with the position given by 4FGL-DR4 source list, and the extended $\gamma$-Cygni source is described as a uniform disk with 68$\%$ error radius $R_{68}$ = 0.517$^{\circ}\!$. Since \citet{2016ApJ...826...31F} and \citet{2023A&A...670A...8M} argue that there are two different spectrum components inside the $\gamma$-Cygni region, here we directly started with two-component templates to fit the $\gamma$-ray emission. First, we used a point-like source plus a uniform disk and 2D-Gaussian template as Model 2 and Model 3 (listed in Table \ref{tab:1}). Their TS values are calculated to be 839 and 846, respectively, both showing a significant improvement compared to the single disk template given by 4FGL-DR4 catalog (Model 1). 

\begin{figure*}
    \centering
    \includegraphics[trim={0 0.cm 0 0}, clip,width=0.32\textwidth]{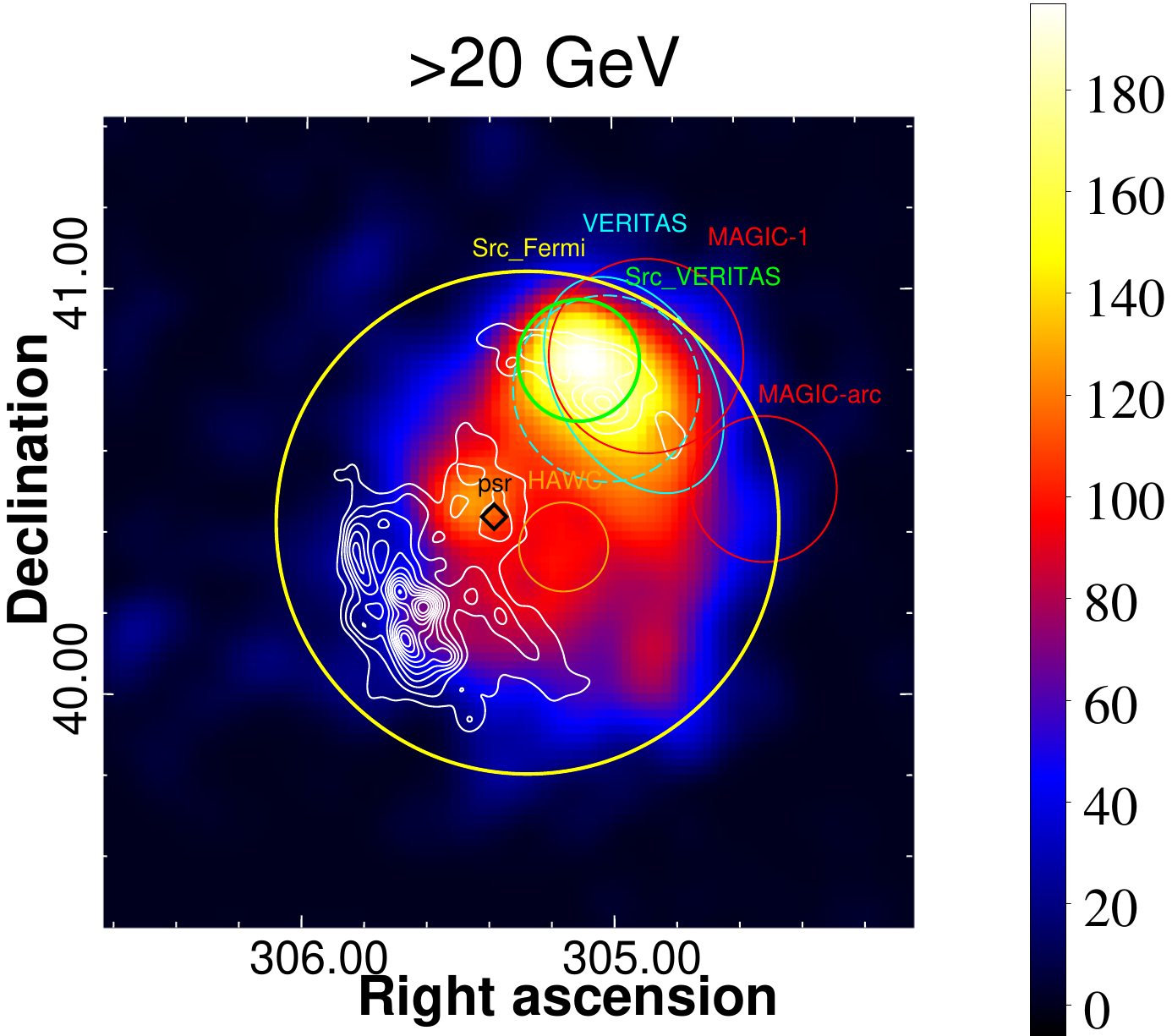}
    \includegraphics[trim={0 0.cm 0 0}, clip,width=0.32\textwidth]{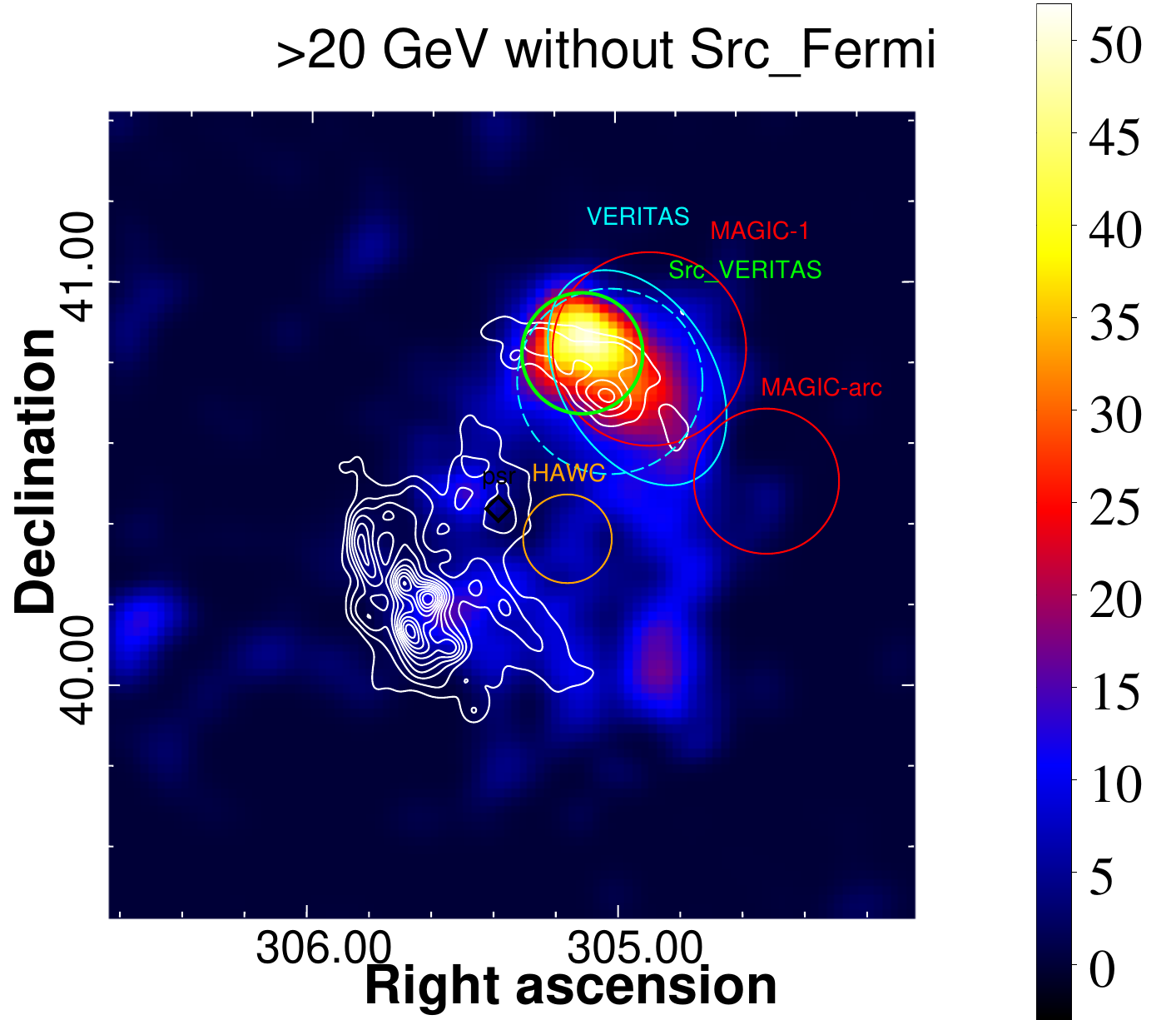}
    \includegraphics[trim={0 0.cm 0 0}, clip,width=0.32\textwidth]{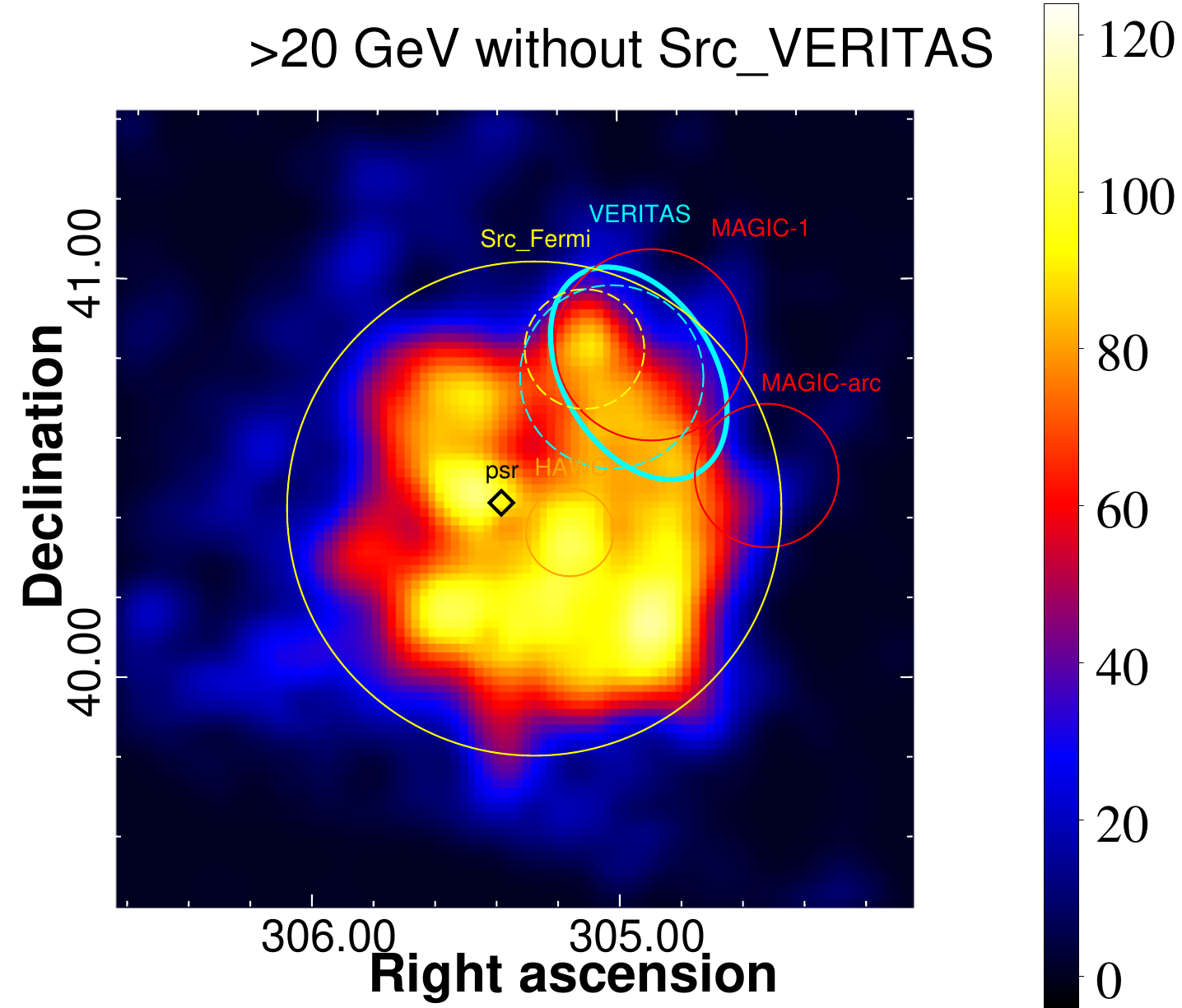} \\
    \includegraphics[trim={0 0.cm 0 0}, clip,width=0.32\textwidth]{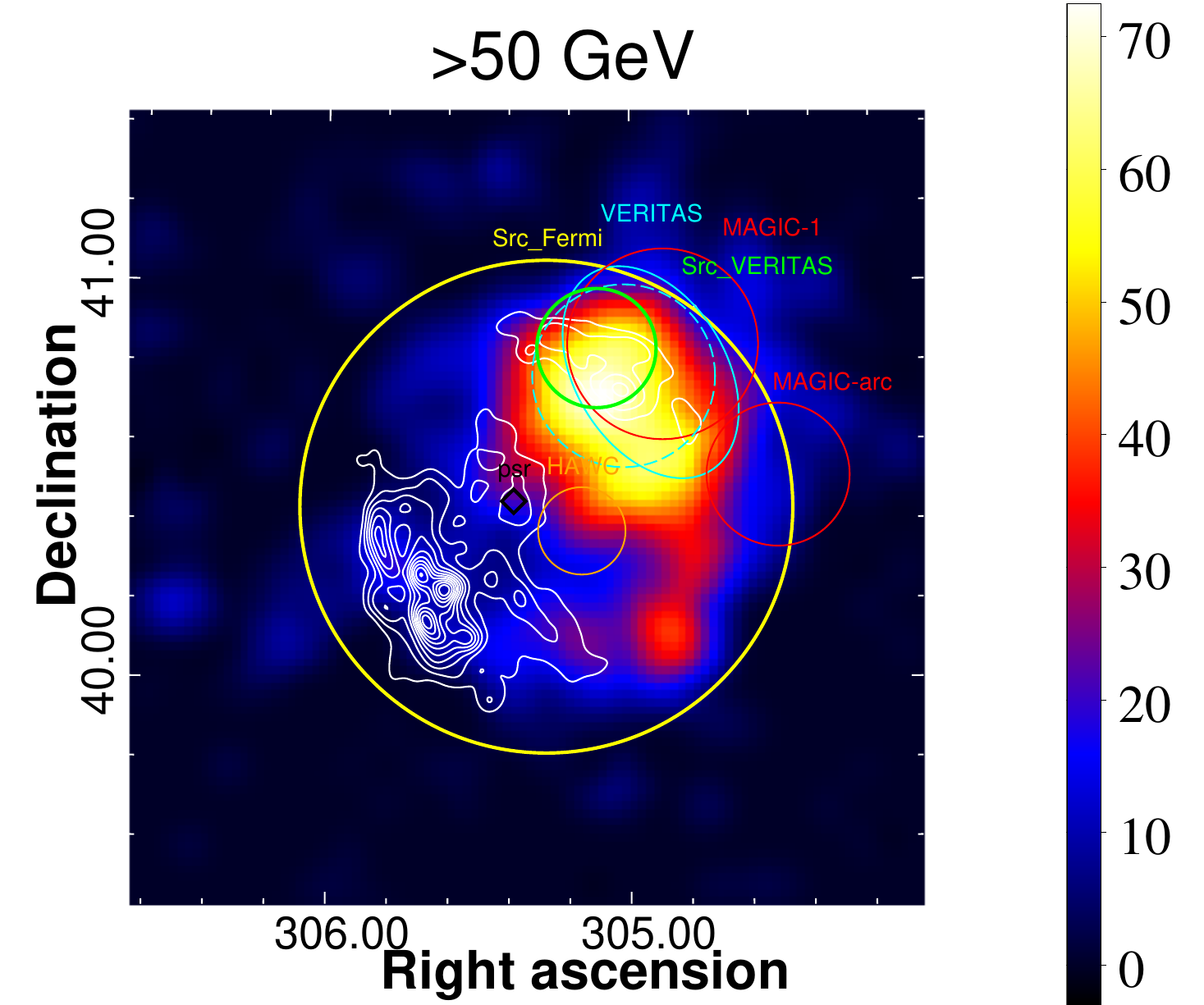}
    \includegraphics[trim={0 0.cm 0 0}, clip,width=0.32\textwidth]{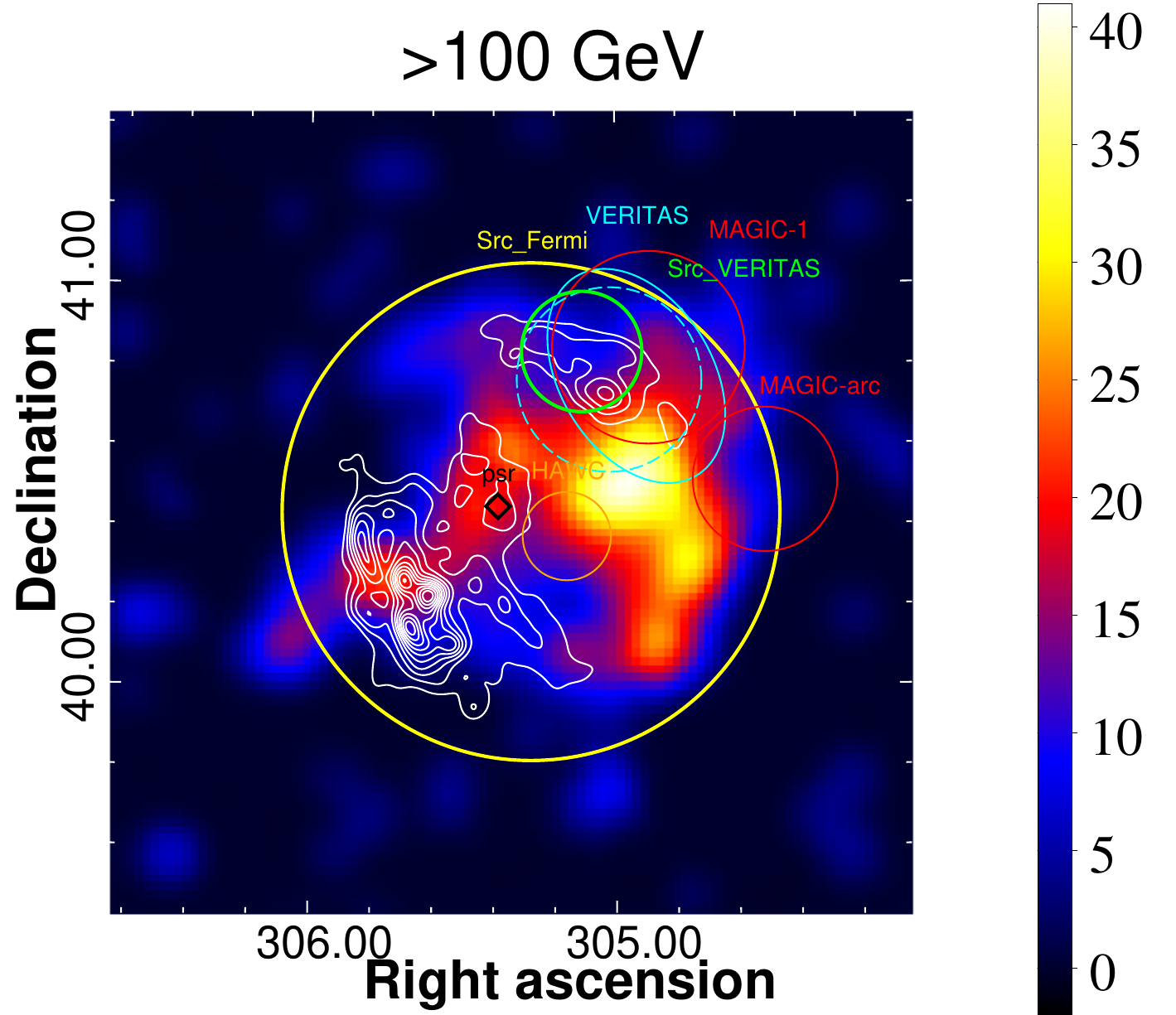}
    \includegraphics[trim={0 0.cm 0 0}, clip,width=0.32\textwidth]{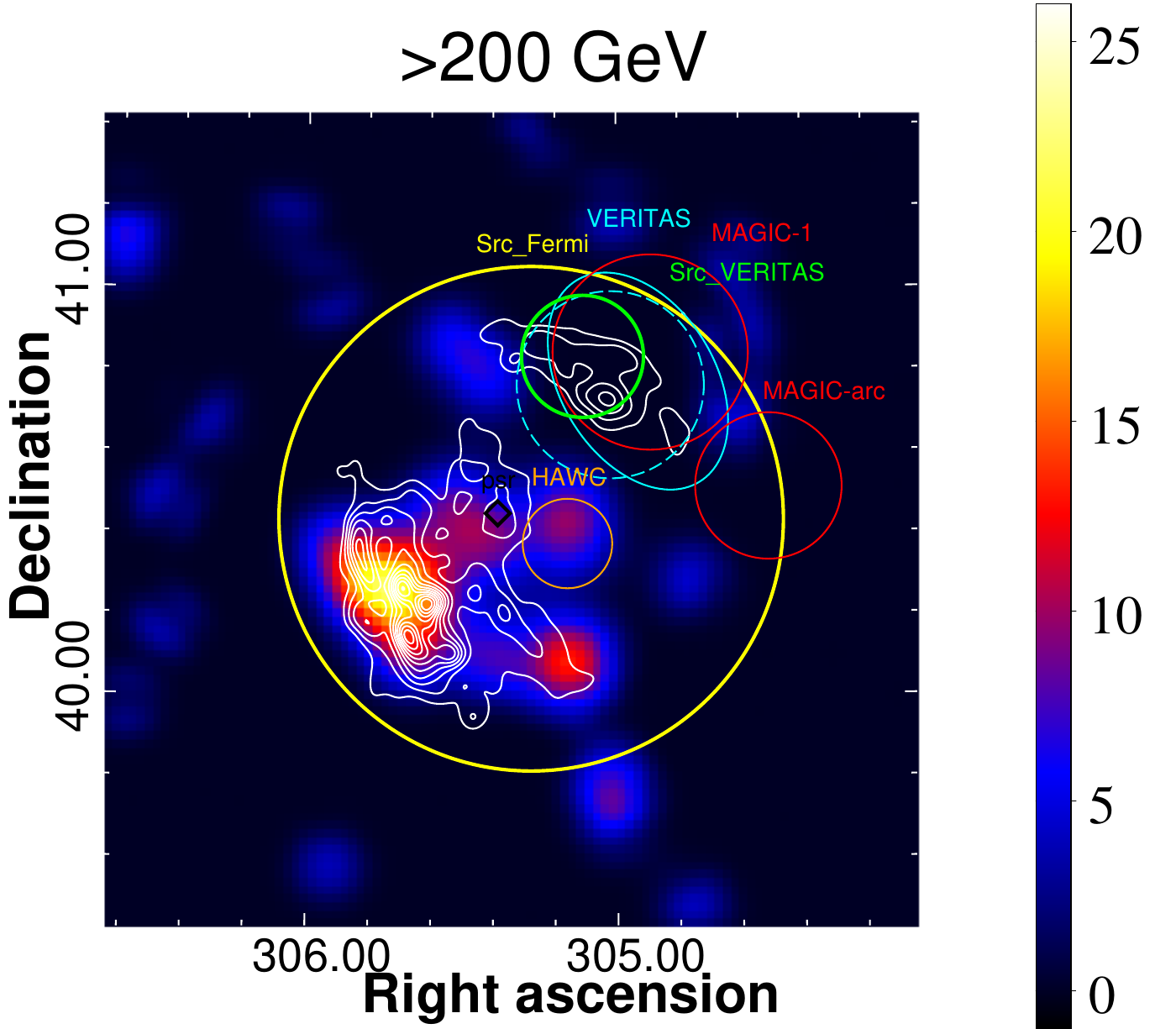}\\
    \caption{\emph{Fermi}-LAT TS maps in the vicinity of $\gamma-$Cygni. The energy range for each sub-figure is shown above them. The yellow and green solid circles show the best-fit two Gaussian template extension size ($R_{68}$) in this work. The red circles show the two extended TeV sources measured by MAGIC \citep{2023A&A...670A...8M}, named as MAGIC-1 and MAGIC-arc, respectively. The cyan circle and ellipse show the VERITAS measured results \citep{2013ApJ...770...93A,2018ApJ...861..134A}, respectively. The orange circle shows the center of HAWC source with 1 $\sigma$ error radius \citep{2017ApJ...843...40A}. The black diamond shows the position of gamma-ray pulsar PSR J2021+4026 \citep{2009Sci...325..840A} same as Figure \ref{fig:9}. The white contours come from the CGPS 1420 MHz radio observation represent the SNR shell structure \citep{2008A&A...490..197L}.} %. Especially, in the top left panel, in order to eliminate the strong pollution from central pulsar, we only used off-pulse phase events.}
   \label{fig:1}
 \end{figure*}

Then we further tested the two extended Gaussian template as Model 4, where the location of the smaller Gaussian is perfectly consistent with TeV observation results given by the VERITAS Collaboration \citep{2013ApJ...770...93A}, and the best-fit results for these two Gaussian sources are recorded as R.A. = $305.277^{\circ}\!$, Dec. = $40.431^{\circ}\!$,  r$_{\rm 68}$ = $0.622^{\circ}\!$ and R.A. = $305.113^{\circ}\!$, Dec. = $40.832^{\circ}\!$,  r$_{\rm 68}$ = $0.151^{\circ}\!$, shown as the yellow and green solid circles in Figure \ref{fig:1} (Hereafter, the 0.622$\degr$ Gaussian source, spatially corresponding to the whole SNR region, is named as Src$_{\rm{Fermi}}$. The 0.151$\degr$ Gaussian source is spatially corresponding to the VERITAS region, and is named as Src$_{\rm{VERITAS}}$, respectively). In this scenario, the $\rm TS_{ext}$ for Src$_{\rm VERITAS}$ is calculated to be approximately 26, rejecting the point-like source hypothesis around the $5.1 \sigma$ level. According to \citet{lande2012}, when $\rm TS_{ext}$ $>$ 16, the extended source hypothesis is valid. These results suggest that the $\gamma$-ray emission from the $\gamma$-Cygni SNR can be described by two extended components. 

To search for the correlation between GeV and TeV emission, we further tested others templates that also include two extended components but directly adopted the location and extension parameters from the VERITAS measurement results. They are labeled as Model 5 and Model 6, and represented by the best-fit 0.622$^\circ$ Gaussian template plus 0.23$^\circ$ circle given by \cite{2013ApJ...770...93A} or $0.29^\circ \times 0.19^\circ$ ellipse given by \cite{2018ApJ...861..134A}, shown as a cyan dashed circle and a cyan solid ellipse in Figure \ref{fig:1}, respectively (Here the uniform ellipse template is generated by the \emph{pyFits}\footnote{\url{ https://pyfits.readthedocs.io/en/latest/index.html}}). Then we tested the 0.622$^\circ$ Gaussian template plus two extended components presented by MAGIC \citep{2023A&A...670A...8M}, labeled as Model 7, shown as red circles in Figure \ref{fig:1}. However, the TS value obtained by Model 5, Model 6 and Model 7 are lower than the two Gaussian template (Model 4). Lastly, to test if there are still others potential sources missed, we additionally added an extended source into Model 4 to fit three Gaussian components simultaneously, the extension and location are all set as free parameters, but the TS value did not increase (Model 8). The results are summarized in Table \ref{tab:1}. In that table, the Akaike information criterion (AIC) value is defined as AIC$ = 2k - 2\ln\mathcal{L}$ and described in \citet{1974AIC}, where $k$ is the number of degrees of freedom of the model and $\mathcal{L}$ is the likelihood value. The model with the minimum AIC value is preferred, thus we adopted Model 4 for the following spectral analysis.

\begin{table*}
\centering
\caption{Spatial models tested for the GeV $\gamma$-ray emission above 20 GeV}
\begin{tabular}{cccccc}
\hline \hline
Morphology($>$20GeV)$^{\,\,\text{a}}$ & TS$^{\,\,\text{b}}$ & R$_{68}$ for $\rm{Src}_{\rm{Fermi}}$&R$_{68}$ for $\rm{Src}_{\rm{VERITAS}}$&Ndf$^{\,\,\text{c}}$&$\Delta${AIC}$^{\,\,\text{d}}$\\
\hline
Model 1 (4FGL-DR4)&770&0.517$\degr$&$-$&5&0 \\
Model 2 (Disk + point)&839&0.471$\degr$&$-$&9&-61 \\
Model 3 (Gaussian + point)&846&0.578$\degr$&$-$&9&-68 \\
Model 4 (Two Gaussians)&872&0.622$\degr$&0.151$\degr$&10&-92 \\
Model 5 (Gaussian + VERITAS circle fixed)&859&0.622$\degr$&0.23$\degr$&4&-91 \\
Model 6 (Gaussian + VERITAS ellipse fixed)&857&0.622$\degr$&(0.29$\degr \times 0.19\degr$)$^{\,\,\text{e}}$&4&-89 \\
Model 7 (Gaussian + MAGIC fixed)&847&0.622$\degr$&0.24$\degr$, 0.18$\degr$&4&-79 \\
Model 8 (Three Gaussians)&849&0.630$\degr$& 0.174$\degr$, 0.213$\degr$ &15&-59 \\
\hline
\end{tabular}
\label{tab:1}
\\
{{\bf Notes.} $^{(a)}$ \textquotesingle Fixed\textquotesingle  in this column represents that the spatial extension parameters were fixed to the TeV source measurement, as mentioned in Section \ref{sec:2.2}. $^{(b)}$ For the global model rather than individual sources}. $^{(c)}$ Degrees of freedom. $^{(d)}$ Calculated with respect to Model 1. $^{(e)}$ Major-axis $\times$ Minor-axis for ellipse.
\end{table*}

\subsection{Energy Spectrum}

\begin{figure*}
    \centering
    \includegraphics[trim={0 0.cm 0 0}, clip, width=0.45\textwidth]{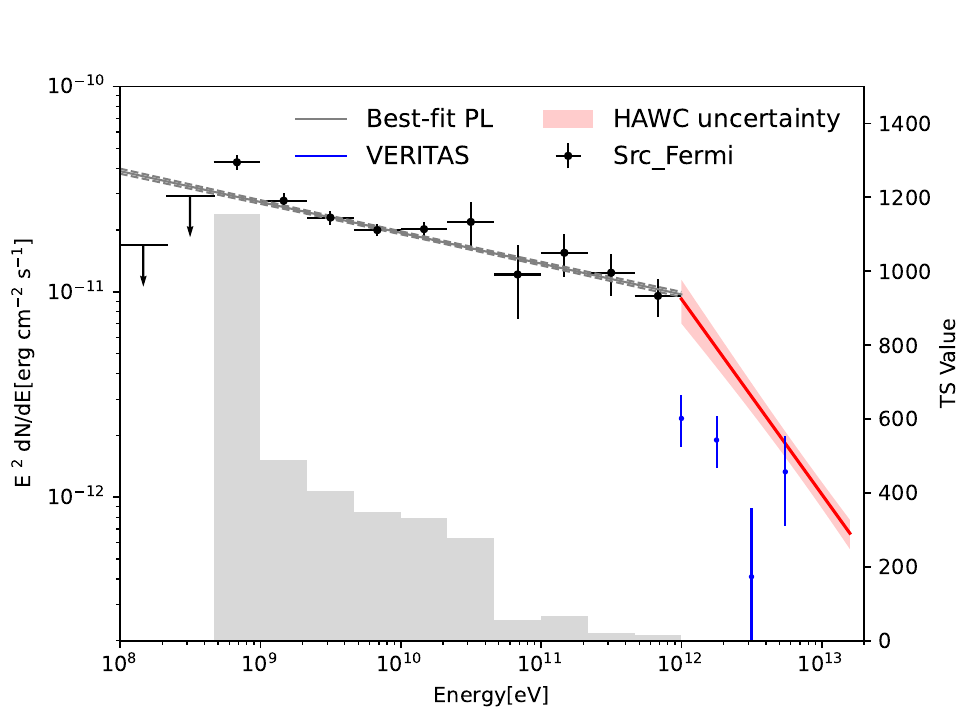}
    \includegraphics[trim={0 0.cm 0 0}, clip, width=0.45\textwidth]{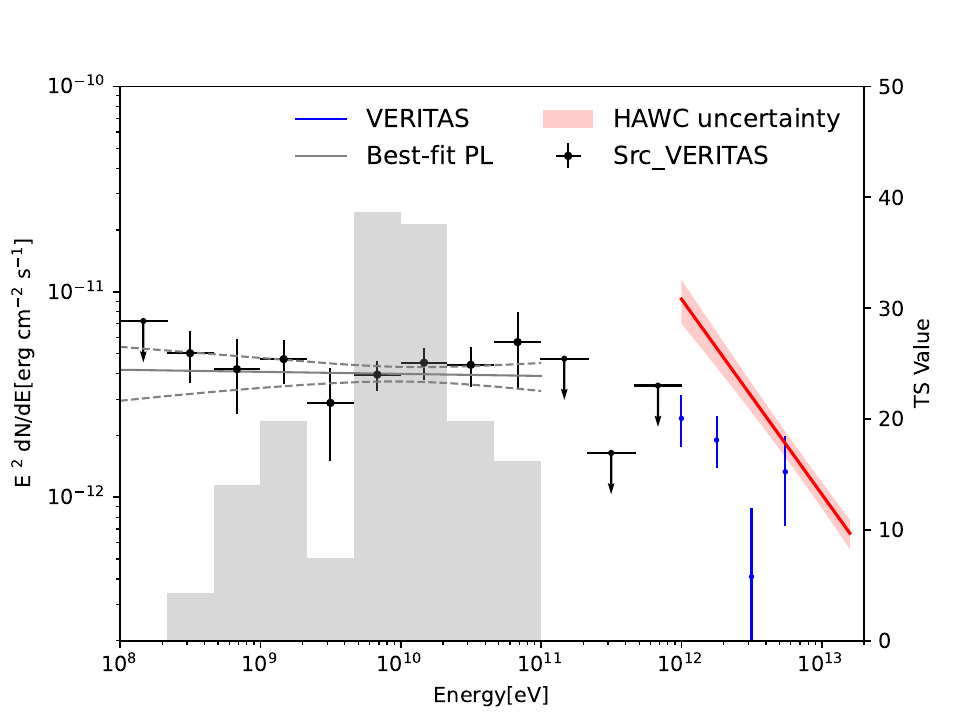}
    \caption{$\gamma$-ray SED obtained from the events in 0.1-1000 GeV shown as black data points. The gray solid line shows the best-fit PL spectra for each component. The dashed lines show the 68$\%$ confidence bands. The arrows indicate the 95$\%$ upper limits when the TS value is less than 4 in a given bin. The blue data points are the VERITAS results \citep{2018ApJ...861..134A}. The red butterfly is extracted from \citet{2017ApJ...843...40A}. The gray histogram denotes the TS value for each bin.}
    \label{fig:2}
\end{figure*}

In Section \ref{sec:2.2}, Model 4 has the best performance and it is adopted here for further spectral analyses. After using events from 100 MeV to 1 TeV, we obtained different spectral types like simple power-law (PL; dN/dE $\propto$ E$^{-\alpha}$) and broken power-law (BPL \footnote{$dN/dE\propto
        \begin{cases}
        \left(\frac{E}{E_b}\right)^{-\Gamma_1}, & E < E_{b} \\
        \left(\frac{E}{E_b}\right)^{-\Gamma_2}, & E > E_{b} \\
        \end{cases}$}
), and the results are then summarized in Table \ref{table:2}. For the Src$_{\rm{Fermi}}$, the BPL model does not provide a notable improvement of the fitting quality compared with a PL assumption. This could be quantified as $\rm{TS_{curve}}$, defined as $\rm{TS_{curve}}=2(\ln\mathcal{L}_{BPL}-\ln\mathcal{L}_{PL})$\citep{abdollahi2020a}, and the derived value of 4.1 corresponds to a significance level of $\sim$ 2.0 $\sigma$. Thus we suggest that there is no energy break in the Src$_{\rm{Fermi}}$ spectrum. For the Src$_{\rm{VERITAS}}$, its $\rm{TS_{curve}}=3.6$ is even lower than 2 $\sigma$ and its calculated TS value shows barely any difference between a PL and a BPL assumption. Thus we suggest that its spectrum can also be described by a single PL. Then we obtained their spectral energy distributions (SEDs) by separating the events in the 100 MeV - 1 TeV energy range into twelve logarithmically equal intervals and repeated the likelihood evaluation analysis for each interval. In this part, the normalization values of all sources are left free, and others parameters are fixed. For bins having TS values lower than 5.0, we calculated the upper limits using a Bayesian method \citep{helene1983} at a 95$\%$ confidence level. The derived SEDs are shown in Fig. \ref{fig:2}, together with the global best-fit spectra in the energy range of 100 MeV - 1 TeV.

\begin{table*}
\caption{Spectral fit parameters in 100 MeV - 1 TeV}
\centering
\begin{tabular}{cccccccc}
\hline
\hline
Sources&Spectral type  &\textbf{$\Gamma$}1&\textbf{$\Gamma$}2&E$_{\rm b}$(GeV)&Photon flux(photon cm$^{-2}$ s$^{-1}$)&TS \\
\hline
Src$_{\rm Fermi}$&PL&2.149 $\pm$ 0.005&$-$&$-$&(3.31 $\pm$ 0.08)$\times$10$^{-7}$&8670 \\
%\hline
Src$_{\rm Fermi}$&BPL&1.683 $\pm$ 0.008&2.171 $\pm$ 0.006&1.6 $\pm$ 0.7&(3.22 $\pm$ 0.07)$\times$10$^{-7}$&8674 \\

Src$_{\rm VERITAS}$&PL&2.01 $\pm$ 0.06&$-$&$-$&(2.34 $\pm$ 0.42)$\times$10$^{-8}$&223 \\

Src$_{\rm VERITAS}$&BPL&1.94 $\pm$ 0.11&2.02 $\pm$ 0.08&3.4 $\pm$ 2.9&(2.41 $\pm$ 0.65)$\times$10$^{-8}$&226 \\
\hline
\hline
\end{tabular}
\label{table:2}
\end{table*}

\section{Gas observations}\label{sec:3}

We make use of the data from the Milky Way Imaging Scroll Painting (MWISP \footnote{\url{http://english.dlh.pmo.cas.cn/ic/}}) project with high resolution carbon monoxide (CO) survey along the Northern Galactic plane with the Purple Mountain Observatory (PMO) 13.7 m telescope \citep{2019ApJS..240....9S}. Furthermore, \citet{2008A&A...490..197L} argued the HI emission at velocities between $\left[-19.0, -11.0\right]$ km s$^{-1}$ and $\left[3.0, 10.0\right]$ km s$^{-1}$ measured by the Canadian Galactic Plane Survey (CGPS) appear to circumscribe the radio continuum emission, which is considered a signature of association with $\gamma$-Cygni, and the average HI spectrum (Fig. 17 in \citet{2008A&A...490..197L}) shows %the symbol (two velocity components) 
that the cloud might be disturbed by the shock front in the velocity range $\left[-16.0, +16.0\right]$ km s$^{-1}$ (see also Figure 37, 38 and 39 in \citet{2008A&A...490..197L}). In the direction of $\gamma$-Cygni, the standard velocity–distance relation becomes unreliable due to complex gas kinematics and significant non-circular motions. As discussed in \citet{2009ApJ...700..137R,2014ApJ...783..130R}, the presence of the Galactic bar, spiral arm shocks, and streaming motions near star-forming regions introduce large deviations from the standard Galactic rotation curve. In particular, the line of sight towards the Cygnus X region ($l \sim 75^\circ$ – $85^\circ$) intersects multiple overlapping molecular cloud complexes \citep{2006A&A...458..855S,2007A&A...474..873S,2012A&A...541A..79G}, making it difficult to associate radial velocity with a unique distance. Moreover, large-scale motions such as super bubble expansions and spiral-arm streaming further distort the local velocity field \citep{2014ApJ...783..130R}. Therefore, instead of relying on a kinematic distance model, we consider the HI studies by \citet{2008A&A...490..197L}, which provide a more accurate representation of the gas structures likely associated with $\gamma$-Cygni in this complex region. Based on this, we integrated the radial velocity range $\left[+9.0, +12.0\right]$ km s$^{-1}$ and $\left[-16.0, -11.0\right]$ km s$^{-1}$ to calculate the molecular hydrogen column mass. 

Here we assume the association between molecular cloud and $\gamma$-Cygni, then adopted the distance \emph{d} = 1.5 kpc into further calculations (which are derived from the assumption that $\gamma$-Cygni is associated with the nearby star-forming region G077.901+01.769 \citep{2012A&A...542A...3S}, while in the Cygnus region direction, we cannot simply obtain the actual distance through velocity). Under the assumption of $\theta$ = $0.622^{\circ}\!$, the obtained physical radius is $\sim$ 16.2 pc. To determine the column density of $\rm H_2$ in this region, we employ a conversion factor $X_\mathrm{CO}=2\times10^{20} \ \rm{cm^{-2} \ K^{-1} \ km^{-1} \ s}$ \citep{bolatto2013,2001ApJ...547..792D}. Using this factor, the column density $N_\mathrm{H_2}$ is calculated as $N_\mathrm{H_2} = X_\mathrm{CO} \times W_\mathrm{CO}$. Consequently, the mass of the molecular complex can be derived from the integrated intensity ($W_\mathrm{CO})$:
\begin{equation}\label{eq:massco}
 M={\mu m_\mathrm{H}} D^2 \Delta\Omega_\mathrm{px} X_\mathrm{CO} {\sum_\mathrm{px}} W_\mathrm{CO} \propto N_\mathrm{H_2},
\end{equation}
In this formula, $\mu$ is set to 2.8, reflecting a relative helium abundance of 25$\%$. The mass of a hydrogen nucleon is denoted as $m_\mathrm{H}$. The solid angle subtended by each pixel is given by $\Delta\Omega_\mathrm{px}$. The term ${\sum_\mathrm{px}} W_\mathrm{CO}$ accounts for the velocity binning of the data cube, which is calculated by summing the map content for the pixels within the target sky region and the specified velocity range, and then scaling by the velocity bin size.

\begin{figure*}
    \centering
    \includegraphics[trim={0 0.cm 0 0}, clip, width=0.45\textwidth]{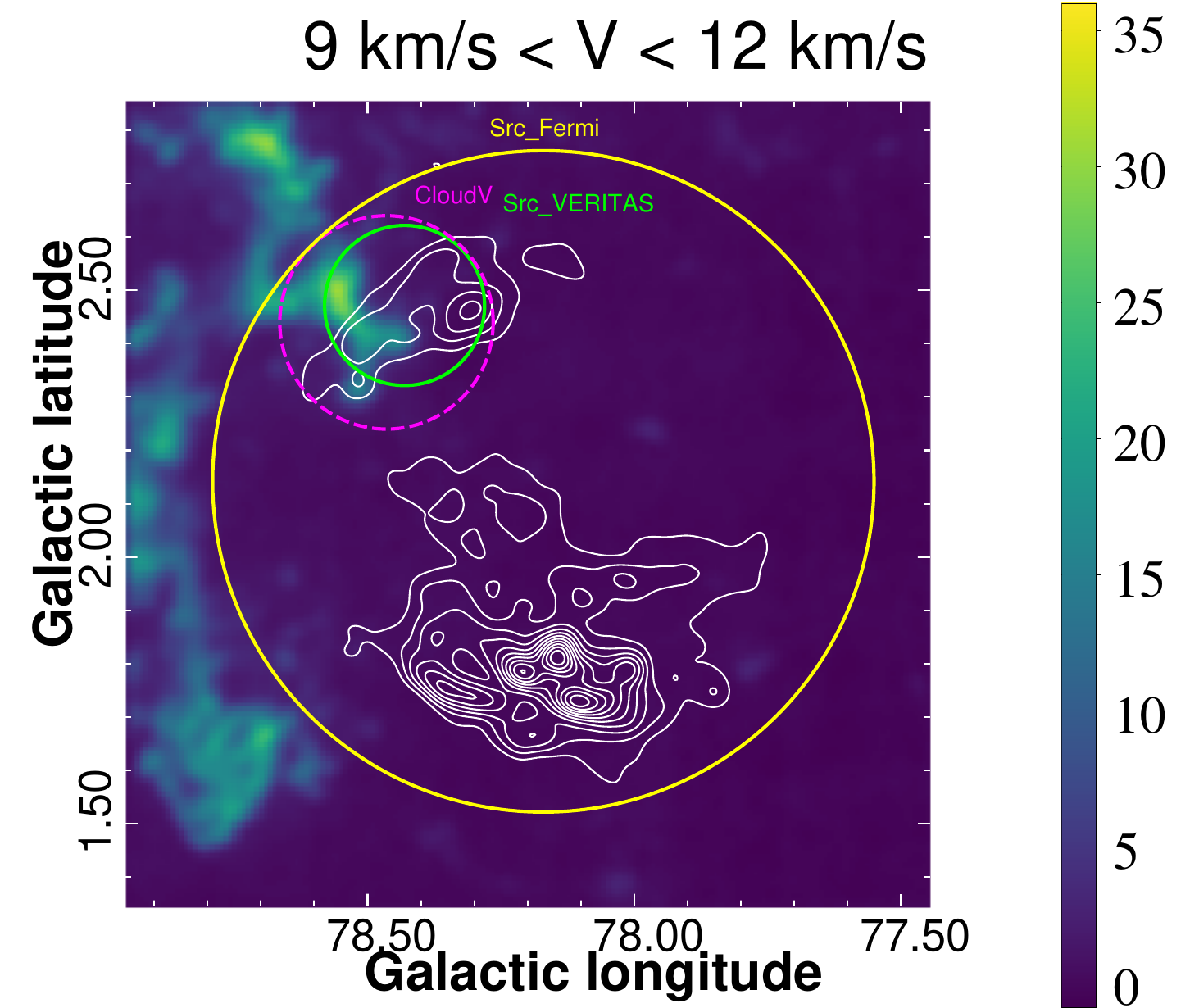}
    \includegraphics[trim={0 0.cm 0 0}, clip, width=0.452\textwidth]{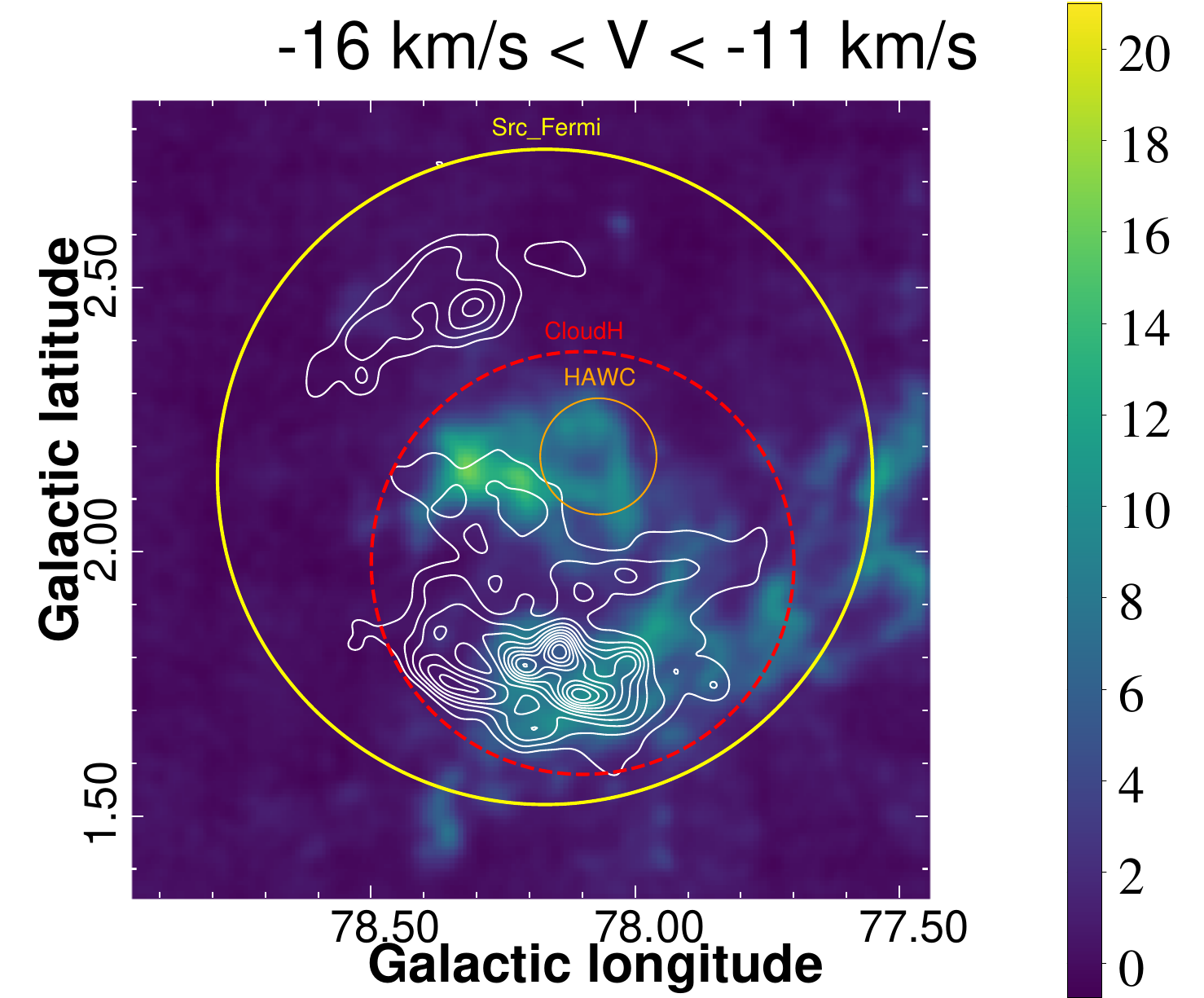}
    \caption{Integrated $^{12}$CO (J = 1-0) emission intensity (K km s$^{-1}$) toward $\gamma$-Cygni region in the velocity ranges $\left[+9.0, +12.0\right]$ km s$^{-1}$ (left panel) and $\left[-16.0, -11.0\right]$ km s$^{-1}$ (right panel) using MWISP data. The solid circles are keep same as Figure \ref{fig:1}. The dashed magenta and dashed red circles correspond to different region affected by escaped CRs, named as CloudV and CloudH, respectively. The $\gamma$-ray contribution within these two spherical regions will be calculated in next section and shown in Fig \ref{fig:4}.}
\label{fig:3}
\end{figure*}

As shown in the left panel of Figure \ref{fig:3}, in the velocity range $\left[+9.0, +12.0\right]$ km s$^{-1}$, only the VERITAS source region exhibits a spatial correspondence between the gas distribution and the location of the $\gamma$-ray emission. The green circle indicates the radius of Src$_{\rm{VERITAS}}$. Since part of the molecular cloud (MC) is located outside the projection size of Src$_{\rm{VERITAS}}$ characterized in \ref{sec:2.2}, the magenta circle is used to calculate the total MC mass in this velocity range, which includes the MC mass potentially interacting with the escaped CRs. In the right panel of Figure \ref{fig:3}, within the velocity range of $\left[-16.0, -11.0\right]$ km s$^{-1}$, the brightest part of the SNR shell in the southeast shows good spatial coincidence with the gas distribution. The red circle denotes the total MC mass in this velocity range, and the orange circle indicates the 1 $\sigma$ uncertainty radius of the HAWC source. Details about their correlation can be found in Sect. \ref{sec:4}. Combined with the HI spectrum, which also supports the association between the SNR and the gas in this velocity range \citep{2023A&A...670A...8M,2008A&A...490..197L}, we calculate the density and mass of each cloud to derive a more precise normalization factor for the subsequent hadronic model.

Hereafter, the dashed magenta and red spherical regions in Figure \ref{fig:3} are named as CloudV and CloudH, respectively, others label in Figure \ref{fig:3} are keep same as Figure \ref{fig:1}. By using the estimation made for $N_\mathrm{H_2}$, in the velocity ranges $\left[+9.0, +12.0\right]$ km s$^{-1}$, the mass of the molecular cloud within Src$_{\rm{VERITAS}}$ region (green circle) is calculated to be about $\rm{M_{veritas}} = 102 \ d_{1.5}^{2} \ M_{\odot}$, the gas mass within magnetic circle region is calculated as $\rm{M_{CloudV}} = 171 \ d_{1.5}^{2} \ M_{\odot}$. Assuming a spherical geometry of the gas distribution, we estimate the volume to be $\rm{V_{veritas}} = {{4\pi \over3}R^3}$, where R = d $\times$ $\theta$, and the average $\rm H_2$ cubic density in this region is about $\rm{n_{veritas}}$ = $\rm 16 \ d_{1.5}^{-1} \ cm^{-3}$. Correspondingly, in the velocity ranges $\left[-16.0, -11.0\right]$ km s$^{-1}$, the gas mass within orange and red spherical regions can be calculated as $\rm{M_{hawc}} = 63 \ d_{1.5}^{2} \ M_{\odot}$ and $\rm{M_{CloudH}} = 542 \ d_{1.5}^{2} \ M_{\odot}$, respectively. Since the $>$100 GeV TS map in Figure \ref{fig:1} shows a decent overlap between the $\gamma$-ray peak and the HAWC source location, with the excess remaining above 3$\sigma$ significance and extending beyond 200 GeV, we assume that the gas within the two source regions (VERITAS and HAWC sources) corresponds to the trapped component, and therefore the gas masses adopted to calculate the $\gamma$-ray fluxes within the trapped emission zones are $\rm{M_{veritas}}$ and $\rm{M_{hawc}}$. Correspondingly, the gas located outside the source region but within the region affected by escaped CRs (CloudV and CloudH shown in Figure \ref{fig:3}) corresponds to the escaped component, which gas mass is adopted to calculate the $\gamma$-ray flux within the escaped emission zone. Therefore we subtract the gas mass in the source region from that in the cloud region to estimate the mass of the escaped component. This yields approximate values of $\rm{M_{escapeH}} = \rm{M_{CloudH}} - \rm{M_{hawc}} = 479 \ d_{1.5}^{2} \ M_{\odot}$ and $\rm{M_{escapeV}} = \rm{M_{CloudV}} - \rm{M_{veritas}} = 69 \ d_{1.5}^{2} \ M_{\odot}$, representing the gas mass associated with the escaped components in different emission zones, respectively. Considering the radius of each spherical region, their densities are calculated to be around several tens of $\rm cm^{-3}$. To be more precise, they fluctuate between approximately 4 $\rm d_{1.5}^{-1} \ cm^{-3}$ and 29 $\rm  d_{1.5}^{-1} \ cm^{-3}$.

\section{Discussion of the possible origins of the $\gamma$-ray emission}\label{sec:4}
\subsection{SNR-MC Interaction}
\begin{figure*}
    \centering
    \includegraphics[trim={0 0.cm 0 0}, clip, width=0.32\textwidth]{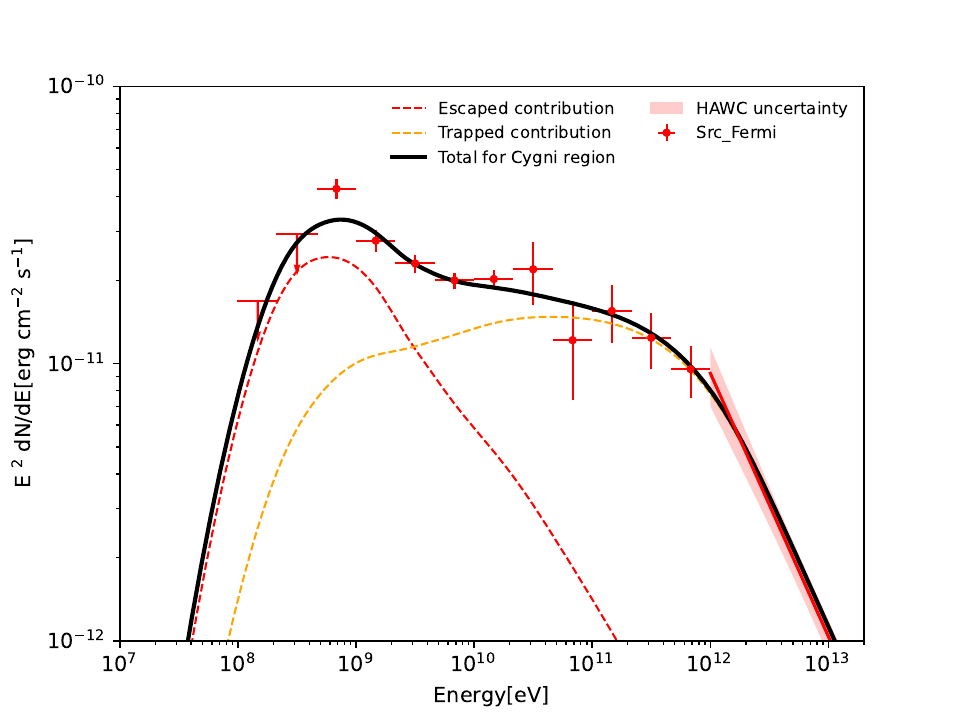}
    \includegraphics[trim={0 0.cm 0 0}, clip, width=0.32\textwidth]{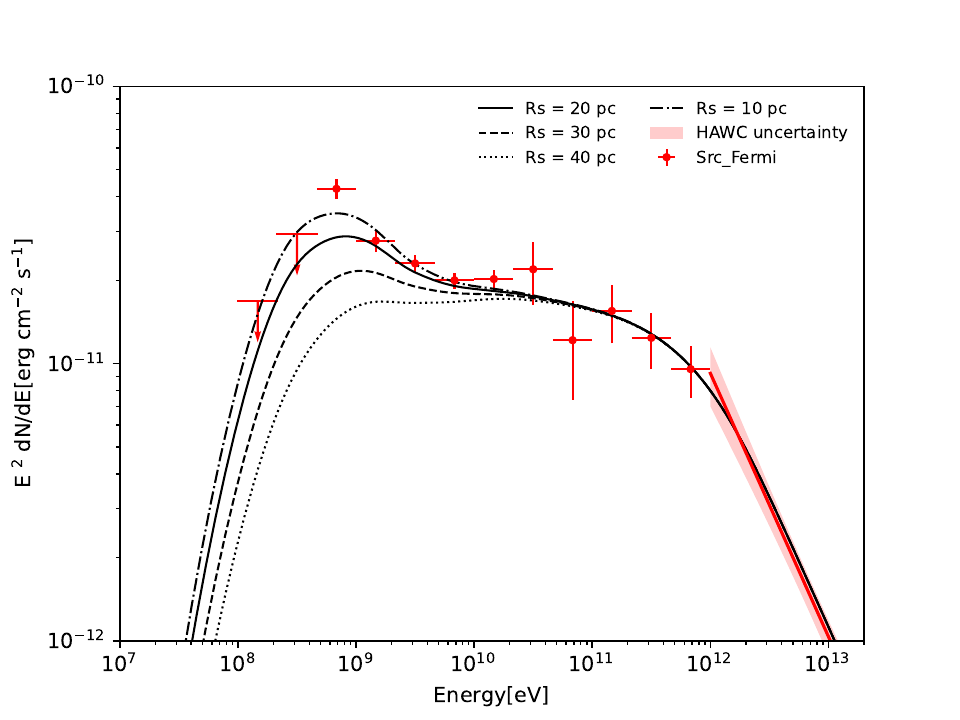}
    \includegraphics[trim={0 0.cm 0 0}, clip, width=0.32\textwidth]{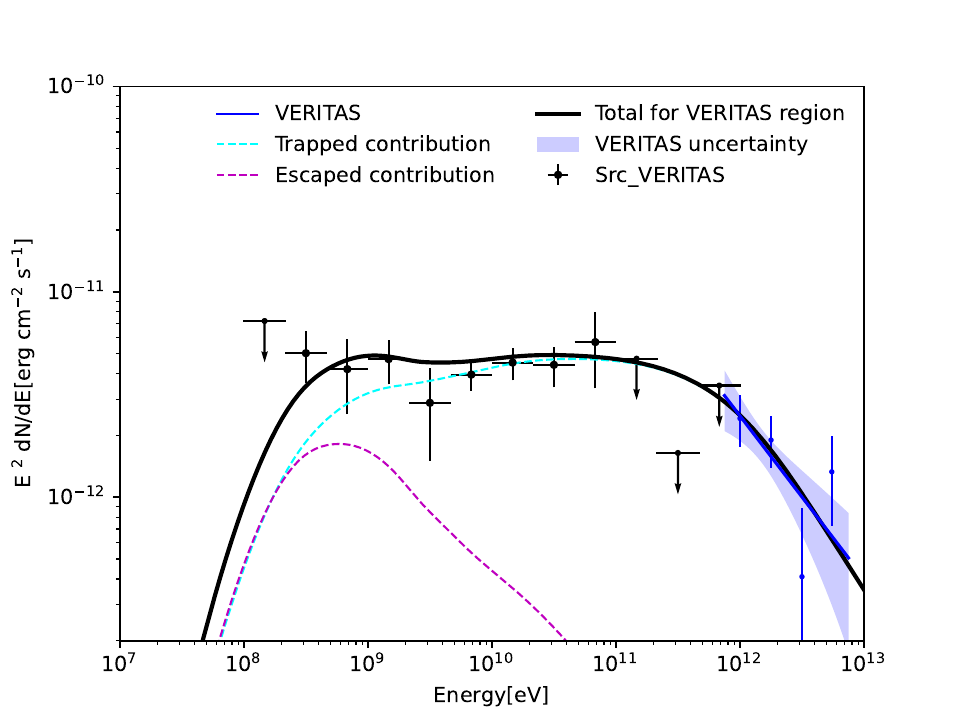}
    \caption{Modeling of the $\gamma$-ray spectrum in the hadronic scenario. Left panel: The black solid line shows the total contribution from the sum of the red and orange dashed lines, corresponding to the escaped and trapped ions, respectively. The red butterfly is extracted from \citet{2017ApJ...843...40A}. Middle panel: Total $\gamma$-ray contribution calculated with different r$_{\rm s}$. Right panel: Predicted $\gamma$-ray flux in the VERITAS region, by adding the escaped and trapped components. Blue data points and butterfly come from \citet{2018ApJ...861..134A}.}
    \label{fig:4}
\end{figure*}

Since the MC distribution shown in Section \ref{sec:3} exhibits good spatial correspondence with both the $\gamma$-ray emission measured by Fermi-LAT and part of the SNR shell, we suggest that the northwestern and southeastern shock fronts are interacting with nearby dense gas, similar to what has been observed in RX J1713.7-3946 \citep{2013ApJ...778...59S,2020ApJ...900L...5T}. Such SNR–MC interactions occurring in lower-density environments tend to result in relatively harder spectral indices in the GeV energy band \citep{2012ApJ...761..133Y}. This scenario implies that only parts of the SNR shell are interacting with molecular clouds of varying densities, leading to inhomogeneous $\gamma$-ray emission. The soft spectral component (escaped contribution shown in Figure \ref{fig:4}) observed in the lower GeV band can be interpreted as escaped cosmic rays (CRs) illuminating nearby MCs, even if their projected positions appear to located within the SNR shell \citep{2023ApJ...953..100L}. In contrast, the flatter spectra component (trapped contribution shown in Figure \ref{fig:4}) is likely produced by CRs still trapped within the shock region and interacting with the local dense gas. Moreover, based on the gas mass calculations presented in Section \ref{sec:3}, the gas mass ratios $\epsilon$ in the two different cloud regions shown in Figure \ref{fig:3} are $\epsilon_{\mathrm{V}} = \rm{M_{\mathrm{veritas}} / M_{\mathrm{escapeV}}} = 102 / 69 \approx 1.48$ and $\epsilon_{\mathrm{H}} = \rm{M_{\mathrm{hawc}} / M_{\mathrm{escapeH}}} = 63 / 479 \approx 0.13$, respectively. These gas mass ratios differ by more than an order of magnitude, indicating that the $\gamma$-ray emission from these two emission regions is mainly contributed by trapped ions and escaped ions, respectively. This difference also naturally accounts for the discrepancy between $\mathrm{Src}_{\mathrm{Fermi}}$ and $\mathrm{Src}_{\mathrm{VERITAS}}$ detections and their distinct spectral characteristics.

Here, we consider a scenario where protons are injected instantaneously into a uniform emission zone (corresponding to CloudV and CloudH mentioned in Section. \ref{sec:3}), approximately 7000 years ago (same as SNR age argued by \citet{2019ApJ...874...50Z}) and the spectrum of the injected protons can be characterized as a broken power-law spectrum:
\begin{eqnarray}
%Q(E) = {Q_0} E^{-\Gamma} \exp \left(- \frac{E}{E_{\rm p, cut}} \right).
Q(E) = {Q_0} \frac{(E/E_{p,\rm break})^{-\gamma_{1}}}{1+(E/E_{p,\rm break})^{\gamma_{2}-\gamma_{1}}} 
\label{eq:p_spectra}
\end{eqnarray}
\\
$E_{p,\rm break}$ represents the break energy, $\gamma_{1}$ and $\gamma_{2}$ being the spectral indices below and above the energy break, respectively. Because of the flat $\gamma$-ray spectrum $\sim$ 2.0 in the GeV energy band for the VERITAS region, we set $\gamma_{1}$ = 2.0 for simplicity and $\gamma_{2}$ = $\gamma_{1}$ + 1.0 \citep{2019ApJ...874...50Z}. $E_{p, \rm break}$ is fixed to 5 TeV, and then the distribution of the escaped protons in the emission region can be calculated following the method outlined by \citet{2020ApJ...897L..34L}:
\begin{equation}
    N_p(E, r_{\rm s}, T)=\frac{Q(E)}{[4 \pi D(E) T]^\frac{3}{2}}\exp\left[\frac{-r_{\rm s}^2}{4 D(E) T}\right] \label{equation:3}
\end{equation}
The diffusion coefficient is assumed to be uniform, following the form $D(E) = \chi D_0 (E/E_0)^\delta$ for $E > E_0$, where $D_0 = 1 \times 10^{28}$ cm$^2$ s$^{-1}$ at $E_0 = 10$ GeV and $\delta = 1/2$, consistent with Kraichnan turbulence \citep{1965PhFl....8.1385K, 2013A&ARv..21...70B}. Due to projection effects, the actual distance between the gas complex and the supernova remnant (SNR) remains uncertain. Furthermore, in the Cygnus region direction, the relationship between kinematic distance and gas velocity can no longer be derived by the method described in \citet{2009ApJ...700..137R,2014ApJ...783..130R}, thus we adopted r$_{\rm s}$ as a free parameter representing the distance between the injection site and the illuminated molecular clouds. With the injected source spectrum defined by $Q(E) \propto E^{-\Gamma}$ and $D(E) \propto E^\delta$, equation \ref{equation:3} reveals that $N_p(E)$ will show a lower energy spectral cutoff at $E_{p, \rm break}$ when $\sqrt{4D(E_b)T} \simeq r_{\rm s}$. Correspondingly, $N_p(E)$ will follow $N_p(E) \propto E^{-\left(\Gamma+\frac{3}{2}\delta\right)}$. The total energy of injected protons is denoted as W$_{\rm inj}$ = $\eta$ E$_{\rm SN}$, where $\eta$ is the efficiency of converting kinetic energy into accelerated protons, typically valued at 0.1. The kinetic energy of the SNR, E$_{\rm SN}$ is generally taken to be 10$^{51}$erg \citep{2013A&ARv..21...70B}. Additionally, the correction factor $\chi$ for the diffusion coefficient is set to 1.0 corresponding to the standard value of the Galactic diffusion coefficient \citep{2013A&ARv..21...70B}. The corresponding $\gamma$-ray fluxes produced in escaped and trapped emission zones are calculated with the $\emph{naima}$ package \citep{zabalza2015naima}.

For the total $\gamma$-Cygni region, the resulting $\gamma$-ray flux can be well described by adding the escaped ions together with the trapped ions contribution, shown as red and orange dashed lines in the left panel of Figure \ref{fig:4}. The total energy of escaped and trapped protons above 1 GeV are calculated to be W$_{\rm escapedH} = 6.32\times 10^{48} (\rm M_{\rm escapeH}/479 M_{\odot})^{-1}$ erg and W$_{\rm trappedH} = 8.98\times 10^{49} (\rm M_{\rm hawc}/63 \rm M_{\odot})^{-1}$ erg, respectively. As mentioned above, assuming a distance of 1.5 kpc, the source radius is calculated to be r$_{\rm s}$ = 16.2 pc. However, the precise value for r$_{\rm s}$ cannot be constrained due to the projection effect, thus we add four scenarios where r$_{\rm s}$ = 10/20/30/40 pc to show the dependence between r$_{\rm s}$ and the $\gamma$-ray flux, and plot the results in the middle panel in Figure \ref{fig:4}. For the VERITAS region, we suggest that its GeV emission with a flat spectrum is mainly dominated by the ongoing shock-cloud interaction \citep{2022RvMPP...6...19L}, shown as a cyan dashed line, while the escaped component also has a contribution. However, due to the relatively high ratio value ($\epsilon_{\rm V}$) of the gas mass between the trapped emission zone and the escaped emission zone, the escaped component can only have a low contribution shown as the magenta dashed line in the right panel of Figure \ref{fig:4}. Correspondingly, the low ratio value ($\epsilon_{\rm H}$) of gas mass between trapped emission zone and escaped emission zone would lead to a higher escaped $\gamma$-ray flux component in the lower energy band, similar to the left panel in Figure \ref{fig:4}. The total energy of each part can be calculated as W$_{\rm escapedV} = 4.77\times 10^{48} (\rm M_{\rm escapeV}/69 \rm M_{\odot})^{-1}$ erg and W$_{\rm trappedV} = 4.31\times 10^{49} (\rm M_{\rm veritas}/102 \rm M_{\odot})^{-1}$ erg. We also calculate the theoretical prediction for the trapped particles in the higher energy band within the Model C proposed by \citet{2015MNRAS.447.2224B}. In this model, the higher energy particles are trapped in a relatively small region compared with the lower energy particles. More specifically, from Figure 5 of \citet{2015MNRAS.447.2224B}, the CRs with energies $\gtrsim$ 100 TeV (corresponding to the $\sim$ 10 TeV photons in HAWC energy band) are predicted to be trapped in a small region inside the SNR, whose radius is roughly 1/5 of the SNR shock radius. The shock radius is $\sim$ 0.5 degree as visible in Figure \ref{fig:3}, therefore this model predicts that the high-energy CRs should be confined in a region with $\sim$ 0.1 degree radius. This is consistent with the measurement from HAWC \citep{2017ApJ...843...40A} that 2HWC J2020+403 is a point-like source, with 1 $\sigma$ uncertainty radius of 0.11 degree. These high-energy CRs correspond to the  CRs that were accelerated early during the expansion phase of the SNR when its shock speed was larger, and that were advected inside the SNR. They remain trapped in tangled magnetic fields from CR-driven instabilities.

\subsection{Pulsar Wind Nebula and Pulsar halo}

Considering the presence of the detected GeV-bright pulsar, the $\gamma$-ray emission in this region might be powered by high-energy electrons and positrons generated by a pulsar wind nebula (PWN). Like, e.g., MSH 15 - 52\citep{2018A&A...612A...2H} and HESS J1825 - 137 \citep{2020A&A...640A..76P}, typical $\gamma$-ray PWNe detected by \emph{Fermi}-LAT are always driven by energetic pulsars with higher $\dot{E}$ between $10^{36} \sim 10^{39}$ erg s$^{-1}$ \citep{2013ApJ...773...77A}. The value of $10^{35}$ erg s$^{-1}$ is at least one order of magnitude lower, which corresponds to the PSR J2021+4026 associated with the $\gamma$-Cygni. Also, the measured photon spectrum indices inside the $\gamma$-Cygni region are much softer than typical PWNe with $\sim$ 1.6 photon index in the GeV energy band, for example MSH 15 - 52\citep{2018A&A...612A...2H} and HESS J1640-465 \citep{2021ApJ...912..158M}. These arguments disfavor the scenario where the $\gamma$-ray emission originates from a PWN. 

\begin{figure*}
    \centering
    \includegraphics[trim={0 0.cm 0 0}, clip, width=0.8\textwidth]{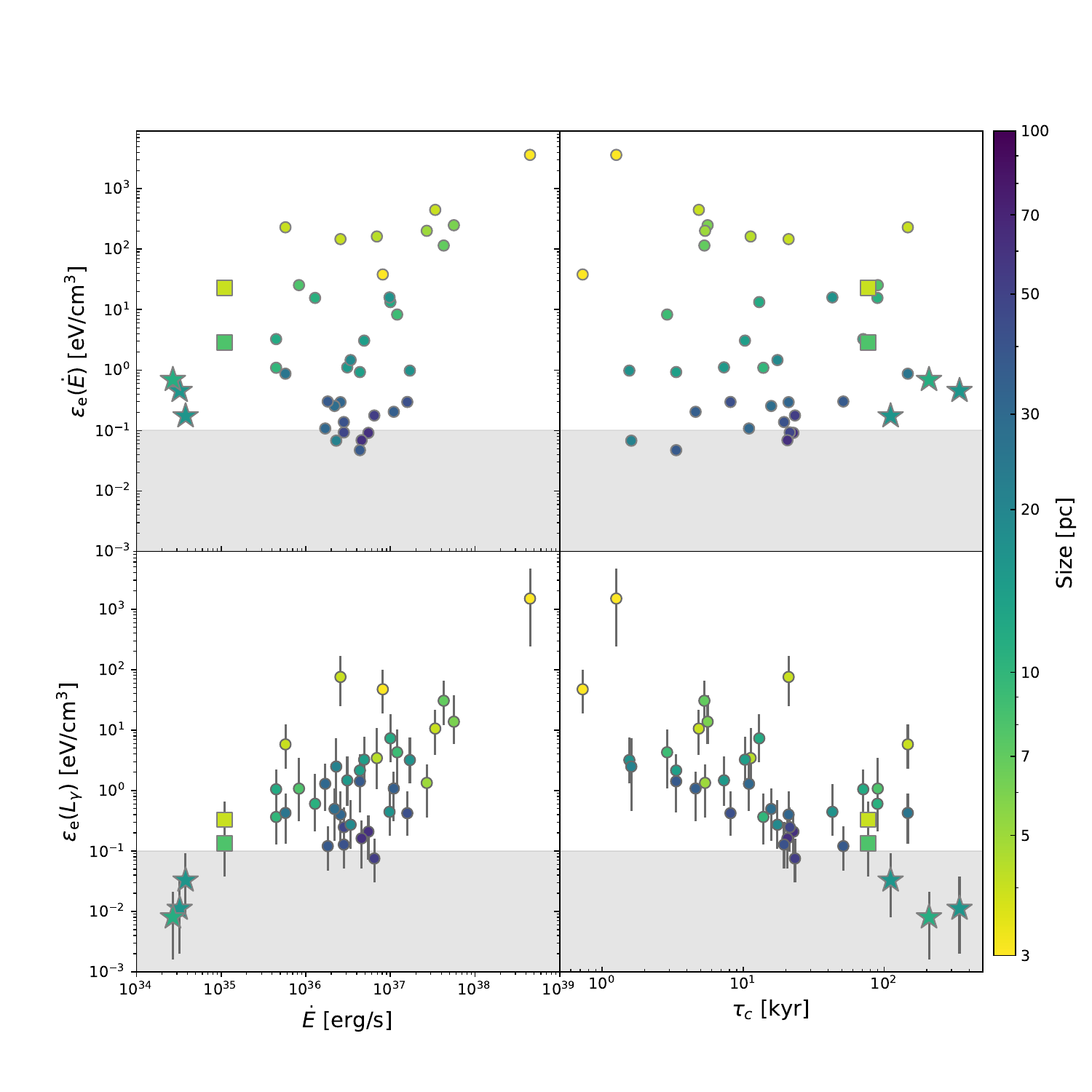}
    \caption{Energy density of TeV sources calculated using two different approaches. The top panels are calculated as $\varepsilon_{\rm{e}} = \dot{E} \tau_c/V$, and the bottom panels are calculated following the approach in Section 3.2 of \citet{2020A&A...636A.113G}. The newly identified pulsar halo associated with PSR J0622+3749 \citep{2021PhRvL.126x1103A}, indicated by a star, is shown along with the Geminga halo and Monogem halo \citep{2017Sci...358..911A}(shown as stars, too). The circular markers representing several $\gamma$-ray PWNe in the figure are based on the data from \citet{2020A&A...636A.113G}. $\gamma$-Cygni under different distance assumptions is shown as squares.}
    \label{fig:5}
\end{figure*}

We now determine if the TeV emission can be interpreted as Inverse Compton scattering of ambient photon fields by relativistic electrons and positrons that have escaped from a potential PWN and diffuse into the ISM \citep{2022NatAs...6..199L}, leading to a pulsar halo around the pulsar. To do so, we adopt here the two different approaches depicted in \citet{2020A&A...636A.113G} to estimate the energy density $\varepsilon_{\rm{e}}$ from the pulsar properties and the $\gamma$-ray luminosity. The region $\varepsilon_{\rm e} \ll \varepsilon_{\rm {ISM}}$ (area shaded in gray in Figure \ref{fig:5}) corresponds to when the energy density in the relativistic electrons and positrons is negligible compared with that of the ISM. This corresponds to the region where pulsar halos may form. In contrast, $\varepsilon_{\rm e} \gtrsim \varepsilon_{\rm {ISM}}$ corresponds to the case when the emitting electrons would be contained in the region that is dynamically dominated by the pulsar \citep{2020A&A...636A.113G}.

To present a more intuitive comparison, we also included the identified pulsar halo LHAASO J0622+3749 \citep{2021PhRvL.126x1103A} in our figure, represented by a star symbol, located near Geminga and PSR B0656+14 \citep{2017Sci...358..911A}. Its parameters are summarized in Table \ref{table:pulsar}. For the $\gamma$-Cygni region, we adopted the TeV extension size of $\theta = 0.23^\circ$ in our calculations. Since the distance to the pulsar is unknown, we considered 1.0 kpc and 2.0 kpc as the minimum and maximum distances, assuming PSR J2021+4026 is associated with the SNR. The corresponding radii for each scenario are 4.014 pc and 8.028 pc, respectively, and they are shown with different squares in the Figure \ref{fig:5}. From the lower panel of Figure \ref{fig:5}, it is hard to definitively attribute the TeV emission to a potential pulsar halo. However, we note that the green square, representing the larger distance (2.0 kpc), is situated near the boundary of shadow region, but not within it. Considering that the pulsar's age is only 77 kyr, we suggest that this source might be in an intermediate stage of transitioning from a PWN to a pulsar halo. This implies that the halo has not yet fully formed, causing its $\dot{E}$ and $\varepsilon_{\rm e}$ to fall between those of a typical PWN and a mature pulsar halo. Additionally, we observe that the calculated $\varepsilon_{\rm{e}}(L_{\gamma})$ is approximately 0.17 eV/cm$^3$ under the 2.0 kpc assumption. This suggests that even if the actual distance is much greater (e.g., 4.0 kpc), the energy density would not decrease sufficiently to fall entirely within the shaded area. These results contrast sharply with those for the three identified pulsar halos shown with stars in Figure \ref{fig:5}, which are all located in the shaded areas in the bottom panels. In the meantime, recent numerical simulations suggest that the morphology of a pulsar halo can exhibit highly asymmetric and distorted shapes, especially when the coherence length of the interstellar turbulence exceeds 10 pc \citep{2018MNRAS.479.4526L}. Furthermore, the simulations of \citet{2024arXiv240702478B} show an offset between the brightest TeV excess and the pulsar position. This implies that, in extreme cases, the actual extension size of the halo may be difficult to measure due to the filamentary structure caused by asymmetric diffusion of cosmic rays \citep{2012PhRvL.108z1101G,2013PhRvD..88b3010G}. Therefore, further investigations of this region using high-resolution detectors across other wavelengths are essential.

\begin{table*}
\centering
\caption {Comparison of the properties of pulsar halo J0622+3749 and $\gamma$-Cygni with different distance assumptions}
\begin{tabular}{ccccccccccc}
\hline \hline
Name & $P$  &$\dot{P}$ & $L_{\rm sd}$ & $\tau$ & $d$ &R$^{\,\,\text{a}}$ & B$^{\,\,\text{b}}$ & $\varepsilon_{\rm{e}}(\dot{E})$ &$\varepsilon_{\rm{e}}(L_{\gamma})$   \\
     & (s)  &($10^{-14}$~s~s$^{-1}$) & ($10^{34}$~erg s$^{-1}$) & (kyr) & (kpc)& (pc) &($10^{12}$ G)&(eV/cm$^3$)&(eV/cm$^3$) \\
\hline
J0622+3749    & $0.333$ & $2.542$  & $2.7$ & 207.8 & 1.60 & 11.17    & 3.0 & 0.644&0.008 \\
Cygni near/far & $0.265$ & $0.348$  & $10.0$ & 77.0 & 1.0/2.0 &4.01/8.02& 4.0& 19.05/2.38&0.34/0.17 \\
\hline
\hline
\end{tabular}
\label{table:pulsar}
{{\bf Notes.} $^{(a)}$ Radius. $^{(b)}$ Magnetic field.}
\end{table*}

\section{Conclusions}\label{sec:5}

We analyzed the $\gamma$-ray emission in the vicinity of $\gamma$-Cygni using 15 years of \emph{Fermi}-LAT data, and confirmed that the GeV counterpart of the VERITAS source is quite different from other parts inside the SNR. Given that the molecular cloud clump in this region spatially coincides with part of the GeV emission and the radio shell, we suggest that the flatter component could be attributed to trapped ions, while the low-energy soft component may be due to escaped cosmic rays interacting with nearby molecular clouds. The GeV spectrum of $\gamma$-Cygni, therefore, consists of a mixture of these two components. The different gas density ratios between the escaped emission zone and the trapped zone result in different $\gamma$-ray flux characteristics across different regions. This scenario matches well the prediction from \citet{2015MNRAS.447.2224B}. In this case, high-energy ions are trapped within a relatively small region around the SNR center, confined by an enhanced magnetic field and unable to escape. The calculated spatial extension of the trapped region agrees well with the observed size of the HAWC source with 1 $\sigma$ uncertainty. When combined with gas observation results, the resulting $\gamma$-ray flux can reasonably explain the measured multi-wavelength data. On the other hand, considering the high spin-down luminosity of the central $\gamma$-ray pulsar PSR J2021+4026, we estimated its energy density $\varepsilon_{\rm{e}}$ using two approaches as shown in \citet{2020A&A...636A.113G}. The calculated energy density shown as the squares in Figure \ref{fig:5} lies well within the range of known PWNe, located near the boundary of known pulsar halos, suggesting that a pulsar halo origin cannot be entirely ruled out. However, the spatial correlation between dense gas, radio SNR shell and $\gamma$-ray emission provides stronger support for the SNR scenario. Taking into account recent numerical simulation work \citep{2018MNRAS.479.4526L,2024arXiv240702478B}, we suggest that the possibility of the SNR overlapping with a twisted-morphology pulsar halo scenario remains plausible.

\begin{acknowledgements}
This research made use of the data from the Milky Way Imaging Scroll Painting (MWISP) project, which is a multi-line survey in 12CO/13CO/C18O along the northern galactic plane with PMO-13.7m telescope. We are grateful to all the members of the MWISP working group, particularly the staff members at PMO-13.7m telescope, for their long-term support. MWISP was sponsored by National Key R$\&$D Program of China with grants 2023YFA1608000 $\&$ 2017YFA0402701 and by CAS Key Research Program of Frontier Sciences with grant QYZDJ-SSW-SLH047. We also would like to thank P.P.Delia, Yang Su for invaluable discussions. This work is supported by the National Natural Science Foundation of China under the grants No. 12393853, U1931204, 12103040, 12147208, and 12350610239, the Natural Science Foundation for Young Scholars of Sichuan Province, China (No. 2022NSFSC1808), and the Fundamental Research Funds for the Central Universities (No. 2682022ZTPY013).
\end{acknowledgements}

\newpage
\bibliographystyle{aa}
\bibliography{ref}

\begin{thebibliography}{68}
\expandafter\ifx\csname natexlab\endcsname\relax\def\natexlab#1{#1}\fi

\bibitem[{{Abdo} {et~al.}(2009{\natexlab{a}}){Abdo}, {Ackermann}, {Ajello},
  {Anderson}, {Atwood}, {Axelsson}, {Baldini}, {Ballet}, {Barbiellini},
  {Baring}, {Bastieri}, {Baughman}, {Bechtol}, {Bellazzini}, {Berenji},
  {Bignami}, {Blandford}, {Bloom}, {Bonamente}, {Borgland}, {Bregeon}, {Brez},
  {Brigida}, {Bruel}, {Burnett}, {Caliandro}, {Cameron}, {Caraveo},
  {Casandjian}, {Cecchi}, {{\c{C}}elik}, {Chekhtman}, {Cheung}, {Chiang},
  {Ciprini}, {Claus}, {Cohen-Tanugi}, {Conrad}, {Cutini}, {Dermer}, {de
  Angelis}, {de Luca}, {de Palma}, {Digel}, {Dormody}, {do Couto e Silva},
  {Drell}, {Dubois}, {Dumora}, {Farnier}, {Favuzzi}, {Fegan}, {Fukazawa},
  {Funk}, {Fusco}, {Gargano}, {Gasparrini}, {Gehrels}, {Germani}, {Giebels},
  {Giglietto}, {Giommi}, {Giordano}, {Glanzman}, {Godfrey}, {Grenier},
  {Grondin}, {Grove}, {Guillemot}, {Guiriec}, {Gwon}, {Hanabata}, {Harding},
  {Hayashida}, {Hays}, {Hughes}, {J{\'o}hannesson}, {Johnson}, {Johnson},
  {Johnson}, {Kamae}, {Katagiri}, {Kataoka}, {Kawai}, {Kerr}, {Kn{\"o}dlseder},
  {Kocian}, {Kuss}, {Lande}, {Latronico}, {Lemoine-Goumard}, {Longo},
  {Loparco}, {Lott}, {Lovellette}, {Lubrano}, {Madejski}, {Makeev}, {Marelli},
  {Mazziotta}, {McConville}, {McEnery}, {Meurer}, {Michelson}, {Mitthumsiri},
  {Mizuno}, {Monte}, {Monzani}, {Morselli}, {Moskalenko}, {Murgia}, {Nolan},
  {Norris}, {Nuss}, {Ohsugi}, {Omodei}, {Orlando}, {Ormes}, {Paneque},
  {Parent}, {Pelassa}, {Pepe}, {Pesce-Rollins}, {Pierbattista}, {Piron},
  {Porter}, {Primack}, {Rain{\`o}}, {Rando}, {Ray}, {Razzano}, {Rea}, {Reimer},
  {Reimer}, {Reposeur}, {Ritz}, {Rochester}, {Rodriguez}, {Romani}, {Ryde},
  {Sadrozinski}, {Sanchez}, {Sander}, {Parkinson}, {Scargle}, {Sgr{\`o}},
  {Siskind}, {Smith}, {Smith}, {Spandre}, {Spinelli}, {Starck}, {Strickman},
  {Suson}, {Tajima}, {Takahashi}, {Takahashi}, {Tanaka}, {Thayer}, {Thompson},
  {Tibaldo}, {Tibolla}, {Torres}, {Tosti}, {Tramacere}, {Uchiyama}, {Usher},
  {Van Etten}, {Vasileiou}, {Vilchez}, {Vitale}, {Waite}, {Wang}, {Watters},
  {Winer}, {Wolff}, {Wood}, {Ylinen}, {Ziegler}, \& {Fermi LAT
  Collaboration}}]{2009Sci...325..840A}
{Abdo}, A.~A., {Ackermann}, M., {Ajello}, M., {et~al.} 2009{\natexlab{a}},
  Science, 325, 840

\bibitem[{{Abdo} {et~al.}(2009{\natexlab{b}}){Abdo}, {Ackermann}, {Ajello},
  {Baldini}, {Ballet}, {Barbiellini}, {Baring}, {Bastieri}, {Baughman},
  {Bechtol}, {Bellazzini}, {Berenji}, {Blandford}, {Bloom}, {Bonamente},
  {Borgland}, {Bouvier}, {Bregeon}, {Brez}, {Brigida}, {Bruel}, {Burnett},
  {Buson}, {Caliandro}, {Cameron}, {Caraveo}, {Casandjian}, {Cecchi},
  {{\c{C}}elik}, {Chekhtman}, {Cheung}, {Chiang}, {Ciprini}, {Claus},
  {Cohen-Tanugi}, {Cominsky}, {Conrad}, {Cutini}, {Dermer}, {de Angelis}, {de
  Palma}, {Digel}, {Dormody}, {Silva}, {Drell}, {Dubois}, {Dumora}, {Farnier},
  {Favuzzi}, {Fegan}, {Focke}, {Fortin}, {Frailis}, {Fukazawa}, {Funk},
  {Fusco}, {Gargano}, {Gasparrini}, {Gehrels}, {Germani}, {Giavitto},
  {Giebels}, {Giglietto}, {Giordano}, {Glanzman}, {Godfrey}, {Grenier},
  {Grondin}, {Grove}, {Guillemot}, {Guiriec}, {Hanabata}, {Harding},
  {Hayashida}, {Hays}, {Hughes}, {Jackson}, {J{\'o}hannesson}, {Johnson},
  {Johnson}, {Johnson}, {Kamae}, {Katagiri}, {Kataoka}, {Katsuta}, {Kawai},
  {Kerr}, {Kn{\"o}dlseder}, {Kocian}, {Kuss}, {Lande}, {Latronico},
  {Lemoine-Goumard}, {Longo}, {Loparco}, {Lott}, {Lovellette}, {Lubrano},
  {Makeev}, {Mazziotta}, {McEnery}, {Meurer}, {Michelson}, {Mitthumsiri},
  {Mizuno}, {Moiseev}, {Monte}, {Monzani}, {Morselli}, {Moskalenko}, {Murgia},
  {Nakamori}, {Nolan}, {Norris}, {Nuss}, {Ohsugi}, {Okumura}, {Omodei},
  {Orlando}, {Ormes}, {Paneque}, {Parent}, {Pelassa}, {Pepe}, {Pesce-Rollins},
  {Piron}, {Porter}, {Rain{\`o}}, {Rando}, {Razzano}, {Reimer}, {Reimer},
  {Reposeur}, {Ritz}, {Rodriguez}, {Romani}, {Roth}, {Ryde}, {Sadrozinski},
  {Sanchez}, {Sander}, {Saz Parkinson}, {Scargle}, {Schalk}, {Sgr{\`o}},
  {Siskind}, {Smith}, {Smith}, {Spandre}, {Spinelli}, {Strickman}, {Suson},
  {Tajima}, {Takahashi}, {Takahashi}, {Tanaka}, {Thayer}, {Thayer}, {Thompson},
  {Tibaldo}, {Tibolla}, {Torres}, {Tosti}, {Tramacere}, {Uchiyama}, {Usher},
  {Vasileiou}, {Venter}, {Vilchez}, {Vitale}, {Waite}, {Wang}, {Winer}, {Wood},
  {Yamazaki}, {Ylinen}, \& {Ziegler}}]{abdo2009fermi}
{Abdo}, A.~A., {Ackermann}, M., {Ajello}, M., {et~al.} 2009{\natexlab{b}},
  \apjl, 706, L1

\bibitem[{{Abdo} {et~al.}(2010){Abdo}, {Ackermann}, {Ajello}, {Baldini},
  {Ballet}, {Barbiellini}, {Bastieri}, {Baughman}, {Bechtol}, {Bellazzini},
  {Berenji}, {Blandford}, {Bloom}, {Bonamente}, {Borgland}, {Bregeon}, {Brez},
  {Brigida}, {Bruel}, {Burnett}, {Buson}, {Caliandro}, {Cameron}, {Caraveo},
  {Casandjian}, {Cecchi}, {{\c{C}}elik}, {Chekhtman}, {Cheung}, {Chiang},
  {Cillis}, {Ciprini}, {Claus}, {Cohen-Tanugi}, {Cominsky}, {Conrad}, {Cutini},
  {Dermer}, {de Angelis}, {de Palma}, {Silva}, {Drell}, {Drlica-Wagner},
  {Dubois}, {Dumora}, {Farnier}, {Favuzzi}, {Fegan}, {Focke}, {Fortin},
  {Frailis}, {Fukazawa}, {Funk}, {Fusco}, {Gargano}, {Gasparrini}, {Gehrels},
  {Germani}, {Giavitto}, {Giebels}, {Giglietto}, {Giordano}, {Glanzman},
  {Godfrey}, {Grenier}, {Grondin}, {Grove}, {Guillemot}, {Guiriec}, {Hanabata},
  {Harding}, {Hayashida}, {Hughes}, {Jackson}, {J{\'o}hannesson}, {Johnson},
  {Johnson}, {Johnson}, {Kamae}, {Katagiri}, {Kataoka}, {Kawai}, {Kerr},
  {Kn{\"o}dlseder}, {Kocian}, {Kuss}, {Lande}, {Latronico}, {Lee},
  {Lemoine-Goumard}, {Longo}, {Loparco}, {Lott}, {Lovellette}, {Lubrano},
  {Madejski}, {Makeev}, {Mazziotta}, {Meurer}, {Michelson}, {Mitthumsiri},
  {Moiseev}, {Monte}, {Monzani}, {Morselli}, {Moskalenko}, {Murgia},
  {Nakamori}, {Nolan}, {Norris}, {Nuss}, {Ohsugi}, {Orlando}, {Ormes}, {Ozaki},
  {Paneque}, {Panetta}, {Parent}, {Pelassa}, {Pepe}, {Pesce-Rollins}, {Piron},
  {Porter}, {Rain{\`o}}, {Rando}, {Razzano}, {Reimer}, {Reimer}, {Reposeur},
  {Rochester}, {Rodriguez}, {Romani}, {Roth}, {Ryde}, {Sadrozinski}, {Sanchez},
  {Sander}, {Saz Parkinson}, {Scargle}, {Sgr{\`o}}, {Siskind}, {Smith},
  {Smith}, {Spandre}, {Spinelli}, {Strickman}, {Strong}, {Suson}, {Tajima},
  {Takahashi}, {Takahashi}, {Tanaka}, {Thayer}, {Thayer}, {Thompson},
  {Tibaldo}, {Torres}, {Tosti}, {Tramacere}, {Uchiyama}, {Usher}, {Van Etten},
  {Vasileiou}, {Venter}, {Vilchez}, {Vitale}, {Waite}, {Wang}, {Winer}, {Wood},
  {Ylinen}, \& {Ziegler}}]{Abdo2010observation}
{Abdo}, A.~A., {Ackermann}, M., {Ajello}, M., {et~al.} 2010, \apj, 712, 459

\bibitem[{{Abdollahi} {et~al.}(2020{\natexlab{a}}){Abdollahi}, {Acero},
  {Ackermann}, {Ajello}, {Atwood}, {Axelsson}, {Baldini}, {Ballet},
  {Barbiellini}, {Bastieri}, {Becerra Gonzalez}, {Bellazzini}, {Berretta},
  {Bissaldi}, {Blandford}, {Bloom}, {Bonino}, {Bottacini}, {Brandt}, {Bregeon},
  {Bruel}, {Buehler}, {Burnett}, {Buson}, {Cameron}, {Caputo}, {Caraveo},
  {Casandjian}, {Castro}, {Cavazzuti}, {Charles}, {Chaty}, {Chen}, {Cheung},
  {Chiaro}, {Ciprini}, {Cohen-Tanugi}, {Cominsky}, {Coronado-Bl{\'a}zquez},
  {Costantin}, {Cuoco}, {Cutini}, {D'Ammando}, {DeKlotz}, {de la Torre Luque},
  {de Palma}, {Desai}, {Digel}, {Di Lalla}, {Di Mauro}, {Di Venere},
  {Dom{\'\i}nguez}, {Dumora}, {Fana Dirirsa}, {Fegan}, {Ferrara},
  {Franckowiak}, {Fukazawa}, {Funk}, {Fusco}, {Gargano}, {Gasparrini},
  {Giglietto}, {Giommi}, {Giordano}, {Giroletti}, {Glanzman}, {Green},
  {Grenier}, {Griffin}, {Grondin}, {Grove}, {Guiriec}, {Harding}, {Hayashi},
  {Hays}, {Hewitt}, {Horan}, {J{\'o}hannesson}, {Johnson}, {Kamae}, {Kerr},
  {Kocevski}, {Kovac'evic'}, {Kuss}, {Landriu}, {Larsson}, {Latronico},
  {Lemoine-Goumard}, {Li}, {Liodakis}, {Longo}, {Loparco}, {Lott},
  {Lovellette}, {Lubrano}, {Madejski}, {Maldera}, {Malyshev}, {Manfreda},
  {Marchesini}, {Marcotulli}, {Mart{\'\i}-Devesa}, {Martin}, {Massaro},
  {Mazziotta}, {McEnery}, {Mereu}, {Meyer}, {Michelson}, {Mirabal}, {Mizuno},
  {Monzani}, {Morselli}, {Moskalenko}, {Negro}, {Nuss}, {Ojha}, {Omodei},
  {Orienti}, {Orlando}, {Ormes}, {Palatiello}, {Paliya}, {Paneque}, {Pei},
  {Pe{\~n}a-Herazo}, {Perkins}, {Persic}, {Pesce-Rollins}, {Petrosian},
  {Petrov}, {Piron}, {Poon}, {Porter}, {Principe}, {Rain{\`o}}, {Rando},
  {Razzano}, {Razzaque}, {Reimer}, {Reimer}, {Remy}, {Reposeur}, {Romani}, {Saz
  Parkinson}, {Schinzel}, {Serini}, {Sgr{\`o}}, {Siskind}, {Smith}, {Spandre},
  {Spinelli}, {Strong}, {Suson}, {Tajima}, {Takahashi}, {Tak}, {Thayer},
  {Thompson}, {Tibaldo}, {Torres}, {Torresi}, {Valverde}, {Van Klaveren}, {van
  Zyl}, {Wood}, {Yassine}, \& {Zaharijas}}]{2020ApJS..247...33A}
{Abdollahi}, S., {Acero}, F., {Ackermann}, M., {et~al.} 2020{\natexlab{a}},
  \apjs, 247, 33

\bibitem[{{Abdollahi} {et~al.}(2020{\natexlab{b}}){Abdollahi}, {Acero},
  {Ackermann}, {Ajello}, {Atwood}, {Axelsson}, {Baldini}, {Ballet},
  {Barbiellini}, {Bastieri}, {Becerra Gonzalez}, {Bellazzini}, {Berretta},
  {Bissaldi}, {Blandford}, {Bloom}, {Bonino}, {Bottacini}, {Brandt}, {Bregeon},
  {Bruel}, {Buehler}, {Burnett}, {Buson}, {Cameron}, {Caputo}, {Caraveo},
  {Casandjian}, {Castro}, {Cavazzuti}, {Charles}, {Chaty}, {Chen}, {Cheung},
  {Chiaro}, {Ciprini}, {Cohen-Tanugi}, {Cominsky}, {Coronado-Bl{\'a}zquez},
  {Costantin}, {Cuoco}, {Cutini}, {D'Ammando}, {DeKlotz}, {de la Torre Luque},
  {de Palma}, {Desai}, {Digel}, {Di Lalla}, {Di Mauro}, {Di Venere},
  {Dom{\'\i}nguez}, {Dumora}, {Fana Dirirsa}, {Fegan}, {Ferrara},
  {Franckowiak}, {Fukazawa}, {Funk}, {Fusco}, {Gargano}, {Gasparrini},
  {Giglietto}, {Giommi}, {Giordano}, {Giroletti}, {Glanzman}, {Green},
  {Grenier}, {Griffin}, {Grondin}, {Grove}, {Guiriec}, {Harding}, {Hayashi},
  {Hays}, {Hewitt}, {Horan}, {J{\'o}hannesson}, {Johnson}, {Kamae}, {Kerr},
  {Kocevski}, {Kovac'evic'}, {Kuss}, {Landriu}, {Larsson}, {Latronico},
  {Lemoine-Goumard}, {Li}, {Liodakis}, {Longo}, {Loparco}, {Lott},
  {Lovellette}, {Lubrano}, {Madejski}, {Maldera}, {Malyshev}, {Manfreda},
  {Marchesini}, {Marcotulli}, {Mart{\'\i}-Devesa}, {Martin}, {Massaro},
  {Mazziotta}, {McEnery}, {Mereu}, {Meyer}, {Michelson}, {Mirabal}, {Mizuno},
  {Monzani}, {Morselli}, {Moskalenko}, {Negro}, {Nuss}, {Ojha}, {Omodei},
  {Orienti}, {Orlando}, {Ormes}, {Palatiello}, {Paliya}, {Paneque}, {Pei},
  {Pe{\~n}a-Herazo}, {Perkins}, {Persic}, {Pesce-Rollins}, {Petrosian},
  {Petrov}, {Piron}, {Poon}, {Porter}, {Principe}, {Rain{\`o}}, {Rando},
  {Razzano}, {Razzaque}, {Reimer}, {Reimer}, {Remy}, {Reposeur}, {Romani}, {Saz
  Parkinson}, {Schinzel}, {Serini}, {Sgr{\`o}}, {Siskind}, {Smith}, {Spandre},
  {Spinelli}, {Strong}, {Suson}, {Tajima}, {Takahashi}, {Tak}, {Thayer},
  {Thompson}, {Tibaldo}, {Torres}, {Torresi}, {Valverde}, {Van Klaveren}, {van
  Zyl}, {Wood}, {Yassine}, \& {Zaharijas}}]{abdollahi2020a}
{Abdollahi}, S., {Acero}, F., {Ackermann}, M., {et~al.} 2020{\natexlab{b}},
  \apjs, 247, 33

\bibitem[{{Abeysekara} {et~al.}(2017{\natexlab{a}}){Abeysekara}, {Albert},
  {Alfaro}, {Alvarez}, {{\'A}lvarez}, {Arceo}, {Arteaga-Vel{\'a}zquez}, {Avila
  Rojas}, {Ayala Solares}, {Barber}, {Bautista-Elivar}, {Becerril},
  {Belmont-Moreno}, {BenZvi}, {Berley}, {Bernal}, {Braun}, {Brisbois},
  {Caballero-Mora}, {Capistr{\'a}n}, {Carrami{\~n}ana}, {Casanova}, {Castillo},
  {Cotti}, {Cotzomi}, {Couti{\~n}o de Le{\'o}n}, {De Le{\'o}n}, {De la Fuente},
  {Dingus}, {DuVernois}, {D{\'\i}az-V{\'e}lez}, {Ellsworth}, {Engel},
  {Enr{\'\i}quez-Rivera}, {Fiorino}, {Fraija}, {Garc{\'\i}a-Gonz{\'a}lez},
  {Garfias}, {Gerhardt}, {Gonz{\'a}lez Mu{\~n}oz}, {Gonz{\'a}lez}, {Goodman},
  {Hampel-Arias}, {Harding}, {Hern{\'a}ndez}, {Hern{\'a}ndez-Almada}, {Hinton},
  {Hona}, {Hui}, {H{\"u}ntemeyer}, {Iriarte}, {Jardin-Blicq}, {Joshi},
  {Kaufmann}, {Kieda}, {Lara}, {Lauer}, {Lee}, {Lennarz}, {Vargas},
  {Linnemann}, {Longinotti}, {Luis Raya}, {Luna-Garc{\'\i}a}, {L{\'o}pez-Coto},
  {Malone}, {Marinelli}, {Martinez}, {Martinez-Castellanos},
  {Mart{\'\i}nez-Castro}, {Mart{\'\i}nez-Huerta}, {Matthews},
  {Miranda-Romagnoli}, {Moreno}, {Mostaf{\'a}}, {Nellen}, {Newbold}, {Nisa},
  {Noriega-Papaqui}, {Pelayo}, {Pretz}, {P{\'e}rez-P{\'e}rez}, {Ren}, {Rho},
  {Rivi{\`e}re}, {Rosa-Gonz{\'a}lez}, {Rosenberg}, {Ruiz-Velasco}, {Salazar},
  {Salesa Greus}, {Sandoval}, {Schneider}, {Schoorlemmer}, {Sinnis}, {Smith},
  {Springer}, {Surajbali}, {Taboada}, {Tibolla}, {Tollefson}, {Torres},
  {Ukwatta}, {Vianello}, {Weisgarber}, {Westerhoff}, {Wisher}, {Wood},
  {Yapici}, {Yodh}, {Younk}, {Zepeda}, {Zhou}, {Guo}, {Hahn}, {Li}, \&
  {Zhang}}]{2017Sci...358..911A}
{Abeysekara}, A.~U., {Albert}, A., {Alfaro}, R., {et~al.} 2017{\natexlab{a}},
  Science, 358, 911

\bibitem[{{Abeysekara} {et~al.}(2017{\natexlab{b}}){Abeysekara}, {Albert},
  {Alfaro}, {Alvarez}, {{\'A}lvarez}, {Arceo}, {Arteaga-Vel{\'a}zquez}, {Ayala
  Solares}, {Barber}, {Baughman}, {Bautista-Elivar}, {Becerra Gonzalez},
  {Becerril}, {Belmont-Moreno}, {BenZvi}, {Berley}, {Bernal}, {Braun},
  {Brisbois}, {Caballero-Mora}, {Capistr{\'a}n}, {Carrami{\~n}ana}, {Casanova},
  {Castillo}, {Cotti}, {Cotzomi}, {Couti{\~n}o de Le{\'o}n}, {de la Fuente},
  {De Le{\'o}n}, {Diaz Hernandez}, {Dingus}, {DuVernois},
  {D{\'\i}az-V{\'e}lez}, {Ellsworth}, {Engel}, {Fiorino}, {Fraija},
  {Garc{\'\i}a-Gonz{\'a}lez}, {Garfias}, {Gerhardt}, {Gonz{\'a}lez Mu{\~n}oz},
  {Gonz{\'a}lez}, {Goodman}, {Hampel-Arias}, {Harding}, {Hernandez},
  {Hernandez-Almada}, {Hinton}, {Hui}, {H{\"u}ntemeyer}, {Iriarte},
  {Jardin-Blicq}, {Joshi}, {Kaufmann}, {Kieda}, {Lara}, {Lauer}, {Lee},
  {Lennarz}, {Le{\'o}n Vargas}, {Linnemann}, {Longinotti}, {Raya},
  {Luna-Garc{\'\i}a}, {L{\'o}pez-Coto}, {Malone}, {Marinelli}, {Martinez},
  {Martinez-Castellanos}, {Mart{\'\i}nez-Castro}, {Mart{\'\i}nez-Huerta},
  {Matthews}, {Miranda-Romagnoli}, {Moreno}, {Mostaf{\'a}}, {Nellen},
  {Newbold}, {Nisa}, {Noriega-Papaqui}, {Pelayo}, {Pretz},
  {P{\'e}rez-P{\'e}rez}, {Ren}, {Rho}, {Rivi{\`e}re}, {Rosa-Gonz{\'a}lez},
  {Rosenberg}, {Ruiz-Velasco}, {Salazar}, {Salesa Greus}, {Sandoval},
  {Schneider}, {Schoorlemmer}, {Sinnis}, {Smith}, {Springer}, {Surajbali},
  {Taboada}, {Tibolla}, {Tollefson}, {Torres}, {Ukwatta}, {Vianello},
  {Villase{\~n}or}, {Weisgarber}, {Westerhoff}, {Wisher}, {Wood}, {Yapici},
  {Younk}, {Zepeda}, \& {Zhou}}]{2017ApJ...843...40A}
{Abeysekara}, A.~U., {Albert}, A., {Alfaro}, R., {et~al.} 2017{\natexlab{b}},
  \apj, 843, 40

\bibitem[{{Abeysekara} {et~al.}(2018){Abeysekara}, {Archer}, {Aune}, {Benbow},
  {Bird}, {Brose}, {Buchovecky}, {Bugaev}, {Cui}, {Daniel}, {Falcone}, {Feng},
  {Finley}, {Fleischhack}, {Flinders}, {Fortson}, {Furniss}, {Gotthelf},
  {Grube}, {Hanna}, {Hervet}, {Holder}, {Huang}, {Hughes}, {Humensky},
  {H{\"u}tten}, {Johnson}, {Kaaret}, {Kar}, {Kelley-Hoskins}, {Kertzman},
  {Kieda}, {Krause}, {Kumar}, {Lang}, {Lin}, {Maier}, {McArthur}, {Moriarty},
  {Mukherjee}, {O'Brien}, {Ong}, {Otte}, {Pandel}, {Park}, {Petrashyk}, {Pohl},
  {Popkow}, {Pueschel}, {Quinn}, {Ragan}, {Reynolds}, {Richards}, {Roache},
  {Rousselle}, {Rulten}, {Sadeh}, {Santander}, {Sembroski}, {Shahinyan},
  {Tyler}, {Vassiliev}, {Wakely}, {Ward}, {Weinstein}, {Wells}, {Wilcox},
  {Wilhelm}, {Williams}, \& {Zitzer}}]{2018ApJ...861..134A}
{Abeysekara}, A.~U., {Archer}, A., {Aune}, T., {et~al.} 2018, \apj, 861, 134

\bibitem[{{Acero} {et~al.}(2013){Acero}, {Ackermann}, {Ajello}, {Allafort},
  {Baldini}, {Ballet}, {Barbiellini}, {Bastieri}, {Bechtol}, {Bellazzini},
  {Blandford}, {Bloom}, {Bonamente}, {Bottacini}, {Brandt}, {Bregeon},
  {Brigida}, {Bruel}, {Buehler}, {Buson}, {Caliandro}, {Cameron}, {Caraveo},
  {Cecchi}, {Charles}, {Chaves}, {Chekhtman}, {Chiang}, {Chiaro}, {Ciprini},
  {Claus}, {Cohen-Tanugi}, {Conrad}, {Cutini}, {Dalton}, {D'Ammando}, {de
  Palma}, {Dermer}, {Di Venere}, {Silva}, {Drell}, {Drlica-Wagner}, {Falletti},
  {Favuzzi}, {Fegan}, {Ferrara}, {Focke}, {Franckowiak}, {Fukazawa}, {Funk},
  {Fusco}, {Gargano}, {Gasparrini}, {Giglietto}, {Giordano}, {Giroletti},
  {Glanzman}, {Godfrey}, {Gr{\'e}goire}, {Grenier}, {Grondin}, {Grove},
  {Guiriec}, {Hadasch}, {Hanabata}, {Harding}, {Hayashida}, {Hayashi}, {Hays},
  {Hewitt}, {Hill}, {Horan}, {Hou}, {Hughes}, {Inoue}, {Jackson}, {Jogler},
  {J{\'o}hannesson}, {Johnson}, {Kamae}, {Kawano}, {Kerr}, {Kn{\"o}dlseder},
  {Kuss}, {Lande}, {Larsson}, {Latronico}, {Lemoine-Goumard}, {Longo},
  {Loparco}, {Lovellette}, {Lubrano}, {Marelli}, {Massaro}, {Mayer},
  {Mazziotta}, {McEnery}, {Mehault}, {Michelson}, {Mitthumsiri}, {Mizuno},
  {Monte}, {Monzani}, {Morselli}, {Moskalenko}, {Murgia}, {Nakamori}, {Nemmen},
  {Nuss}, {Ohsugi}, {Okumura}, {Orienti}, {Orlando}, {Ormes}, {Paneque},
  {Panetta}, {Perkins}, {Pesce-Rollins}, {Piron}, {Pivato}, {Porter},
  {Rain{\`o}}, {Rando}, {Razzano}, {Reimer}, {Reimer}, {Reposeur}, {Ritz},
  {Roth}, {Rousseau}, {Saz Parkinson}, {Schulz}, {Sgr{\`o}}, {Siskind},
  {Smith}, {Spandre}, {Spinelli}, {Suson}, {Takahashi}, {Takeuchi}, {Thayer},
  {Thayer}, {Thompson}, {Tibaldo}, {Tibolla}, {Tinivella}, {Torres}, {Tosti},
  {Troja}, {Uchiyama}, {Vandenbroucke}, {Vasileiou}, {Vianello}, {Vitale},
  {Werner}, {Winer}, {Wood}, \& {Yang}}]{2013ApJ...773...77A}
{Acero}, F., {Ackermann}, M., {Ajello}, M., {et~al.} 2013, \apj, 773, 77

\bibitem[{{Ackermann} {et~al.}(2013){Ackermann}, {Ajello}, {Allafort},
  {Baldini}, {Ballet}, {Barbiellini}, {Baring}, {Bastieri}, {Bechtol},
  {Bellazzini}, {Blandford}, {Bloom}, {Bonamente}, {Borgland}, {Bottacini},
  {Brandt}, {Bregeon}, {Brigida}, {Bruel}, {Buehler}, {Busetto}, {Buson},
  {Caliandro}, {Cameron}, {Caraveo}, {Casandjian}, {Cecchi}, {{\c{C}}elik},
  {Charles}, {Chaty}, {Chaves}, {Chekhtman}, {Cheung}, {Chiang}, {Chiaro},
  {Cillis}, {Ciprini}, {Claus}, {Cohen-Tanugi}, {Cominsky}, {Conrad}, {Corbel},
  {Cutini}, {D'Ammando}, {de Angelis}, {de Palma}, {Dermer}, {do Couto e
  Silva}, {Drell}, {Drlica-Wagner}, {Falletti}, {Favuzzi}, {Ferrara},
  {Franckowiak}, {Fukazawa}, {Funk}, {Fusco}, {Gargano}, {Germani},
  {Giglietto}, {Giommi}, {Giordano}, {Giroletti}, {Glanzman}, {Godfrey},
  {Grenier}, {Grondin}, {Grove}, {Guiriec}, {Hadasch}, {Hanabata}, {Harding},
  {Hayashida}, {Hayashi}, {Hays}, {Hewitt}, {Hill}, {Hughes}, {Jackson},
  {Jogler}, {J{\'o}hannesson}, {Johnson}, {Kamae}, {Kataoka}, {Katsuta},
  {Kn{\"o}dlseder}, {Kuss}, {Lande}, {Larsson}, {Latronico}, {Lemoine-Goumard},
  {Longo}, {Loparco}, {Lovellette}, {Lubrano}, {Madejski}, {Massaro}, {Mayer},
  {Mazziotta}, {McEnery}, {Mehault}, {Michelson}, {Mignani}, {Mitthumsiri},
  {Mizuno}, {Moiseev}, {Monzani}, {Morselli}, {Moskalenko}, {Murgia},
  {Nakamori}, {Nemmen}, {Nuss}, {Ohno}, {Ohsugi}, {Omodei}, {Orienti},
  {Orlando}, {Ormes}, {Paneque}, {Perkins}, {Pesce-Rollins}, {Piron}, {Pivato},
  {Rain{\`o}}, {Rando}, {Razzano}, {Razzaque}, {Reimer}, {Reimer}, {Ritz},
  {Romoli}, {S{\'a}nchez-Conde}, {Schulz}, {Sgr{\`o}}, {Simeon}, {Siskind},
  {Smith}, {Spandre}, {Spinelli}, {Stecker}, {Strong}, {Suson}, {Tajima},
  {Takahashi}, {Takahashi}, {Tanaka}, {Thayer}, {Thayer}, {Thompson},
  {Thorsett}, {Tibaldo}, {Tibolla}, {Tinivella}, {Troja}, {Uchiyama}, {Usher},
  {Vandenbroucke}, {Vasileiou}, {Vianello}, {Vitale}, {Waite}, {Werner},
  {Winer}, {Wood}, {Wood}, {Yamazaki}, {Yang}, \& {Zimmer}}]{Ackermann2013}
{Ackermann}, M., {Ajello}, M., {Allafort}, A., {et~al.} 2013, Science, 339, 807

\bibitem[{{Aharonian} {et~al.}(2008){Aharonian}, {Akhperjanian}, {Bazer-Bachi},
  {Behera}, {Beilicke}, {Benbow}, {Berge}, {Bernl{\"o}hr}, {Boisson}, {Bolz},
  {Borrel}, {Braun}, {Brion}, {Brown}, {B{\"u}hler}, {Bulik}, {B{\"u}sching},
  {Boutelier}, {Carrigan}, {Chadwick}, {Chounet}, {Clapson}, {Coignet},
  {Cornils}, {Costamante}, {Degrange}, {Dickinson}, {Djannati-Ata{\"\i}},
  {Domainko}, {O'C. Drury}, {Dubus}, {Dyks}, {Egberts}, {Emmanoulopoulos},
  {Espigat}, {Farnier}, {Feinstein}, {Fiasson}, {F{\"o}rster}, {Fontaine},
  {Fukui}, {Funk}, {Funk}, {F{\"u}{\ss}ling}, {Gallant}, {Giebels},
  {Glicenstein}, {Gl{\"u}ck}, {Goret}, {Hadjichristidis}, {Hauser}, {Hauser},
  {Heinzelmann}, {Henri}, {Hermann}, {Hinton}, {Hoffmann}, {Hofmann},
  {Holleran}, {Hoppe}, {Horns}, {Jacholkowska}, {de Jager}, {Kendziorra},
  {Kerschhaggl}, {Kh{\'e}lifi}, {Komin}, {Kosack}, {Lamanna}, {Latham}, {Le
  Gallou}, {Lemi{\`e}re}, {Lemoine-Goumard}, {Lenain}, {Lohse}, {Martin},
  {Martineau-Huynh}, {Marcowith}, {Masterson}, {Maurin}, {McComb}, {Moderski},
  {Moriguchi}, {Moulin}, {de Naurois}, {Nedbal}, {Nolan}, {Olive}, {Orford},
  {Osborne}, {Ostrowski}, {Panter}, {Pedaletti}, {Pelletier}, {Petrucci},
  {Pita}, {P{\"u}hlhofer}, {Punch}, {Ranchon}, {Raubenheimer}, {Raue},
  {Rayner}, {Reimer}, {Renaud}, {Ripken}, {Rob}, {Rolland}, {Rosier-Lees},
  {Rowell}, {Rudak}, {Ruppel}, {Sahakian}, {Santangelo}, {Saug{\'e}},
  {Schlenker}, {Schlickeiser}, {Schr{\"o}der}, {Schwanke}, {Schwarzburg},
  {Schwemmer}, {Shalchi}, {Sol}, {Spangler}, {Stawarz}, {Steenkamp},
  {Stegmann}, {Superina}, {Takeuchi}, {Tam}, {Tavernet}, {Terrier}, {van
  Eldik}, {Vasileiadis}, {Venter}, {Vialle}, {Vincent}, {Vivier}, {V{\"o}lk},
  {Volpe}, {Wagner}, \& {Ward}}]{aharonian2008discovery}
{Aharonian}, F., {Akhperjanian}, A.~G., {Bazer-Bachi}, A.~R., {et~al.} 2008,
  \aap, 481, 401

\bibitem[{{Aharonian} {et~al.}(2021){Aharonian}, {An}, {Axikegu}, {Bai}, {Bao},
  {Bastieri}, {Bi}, {Bi}, {Cai}, {Cai}, {Cao}, {Cao}, {Chang}, {Chang},
  {Chang}, {Chen}, {Chen}, {Chen}, {Chen}, {Chen}, {Chen}, {Chen}, {Chen},
  {Chen}, {Chen}, {Chen}, {Chen}, {Chen}, {Cheng}, {Cheng}, {Cui}, {Cui},
  {Cui}, {Dai}, {Dai}, {Dai}, {Danzengluobu}, {Della Volpe}, {D'Ettorre
  Piazzoli}, {Dong}, {Fan}, {Fan}, {Fan}, {Fang}, {Fang}, {Feng}, {Feng},
  {Feng}, {Feng}, {Gao}, {Gao}, {Gao}, {Gao}, {Ge}, {Geng}, {Gong}, {Gou},
  {Gu}, {Guo}, {Guo}, {Guo}, {Guo}, {Han}, {He}, {He}, {He}, {He}, {He}, {He},
  {Heller}, {Hor}, {Hou}, {Hou}, {Hu}, {Hu}, {Hu}, {Hu}, {Huang}, {Huang},
  {Huang}, {Huang}, {Huang}, {Ji}, {Ji}, {Jia}, {Jiang}, {Jiang}, {Jin},
  {Kuleshov}, {Levochkin}, {Li}, {Li}, {Li}, {Li}, {Li}, {Li}, {Li}, {Li},
  {Li}, {Li}, {Li}, {Li}, {Li}, {Li}, {Li}, {Li}, {Li}, {Liang}, {Liang},
  {Lin}, {Liu}, {Liu}, {Liu}, {Liu}, {Liu}, {Liu}, {Liu}, {Liu}, {Liu}, {Liu},
  {Liu}, {Liu}, {Liu}, {Liu}, {Liu}, {Long}, {Lu}, {Lv}, {Ma}, {Ma}, {Ma},
  {Mao}, {Masood}, {Mitthumsiri}, {Montaruli}, {Nan}, {Pang},
  {Pattarakijwanich}, {Pei}, {Qi}, {Ruffolo}, {Rulev}, {S{\'a}iz}, {Shao},
  {Shchegolev}, {Sheng}, {Shi}, {Song}, {Stenkin}, {Stepanov}, {Sun}, {Sun},
  {Sun}, {Tam}, {Tang}, {Tian}, {Wang}, {Wang}, {Wang}, {Wang}, {Wang}, {Wang},
  {Wang}, {Wang}, {Wang}, {Wang}, {Wang}, {Wang}, {Wang}, {Wang}, {Wang},
  {Wang}, {Wang}, {Wang}, {Wang}, {Wang}, {Wang}, {Wei}, {Wei}, {Wei}, {Wen},
  {Wu}, {Wu}, {Wu}, {Wu}, {Wu}, {Xi}, {Xia}, {Xia}, {Xiang}, {Xiao}, {Xiao},
  {Xin}, {Xin}, {Xing}, {Xu}, {Xu}, {Xue}, {Yan}, {Yang}, {Yang}, {Yang},
  {Yang}, {Yang}, {Yang}, {Yang}, {Yao}, {Yao}, {Ye}, {Yin}, {Yin}, {You},
  {You}, {Yu}, {Yuan}, {Zeng}, {Zeng}, {Zeng}, {Zeng}, {Zha}, {Zhai}, {Zhang},
  {Zhang}, {Zhang}, {Zhang}, {Zhang}, {Zhang}, {Zhang}, {Zhang}, {Zhang},
  {Zhang}, {Zhang}, {Zhang}, {Zhang}, {Zhang}, {Zhang}, {Zhang}, {Zhang},
  {Zhang}, {Zhang}, {Zhao}, {Zhao}, {Zhao}, {Zhao}, {Zhao}, {Zheng}, {Zheng},
  {Zhou}, {Zhou}, {Zhou}, {Zhou}, {Zhou}, {Zhou}, {Zhu}, {Zhu}, {Zhu}, {Zhu},
  {Zuo}, {LHAASO Collaboration}, \& {Huang}}]{2021PhRvL.126x1103A}
{Aharonian}, F., {An}, Q., {Axikegu}, Bai, L.~X., {et~al.} 2021, \prl, 126,
  241103

\bibitem[{{Akaike}(1974)}]{1974AIC}
{Akaike}, H. 1974, IEEE Transactions on Automatic Control, 19, 716

\bibitem[{{Aliu} {et~al.}(2013){Aliu}, {Archambault}, {Arlen}, {Aune},
  {Beilicke}, {Benbow}, {Bird}, {Bouvier}, {Bradbury}, {Buckley}, {Bugaev},
  {Byrum}, {Cannon}, {Cesarini}, {Ciupik}, {Collins-Hughes}, {Connolly}, {Cui},
  {Dickherber}, {Duke}, {Dumm}, {Dwarkadas}, {Errando}, {Falcone}, {Federici},
  {Feng}, {Finley}, {Finnegan}, {Fortson}, {Furniss}, {Galante}, {Gall},
  {Gillanders}, {Godambe}, {Gotthelf}, {Griffin}, {Grube}, {Gyuk}, {Hanna},
  {Holder}, {Huan}, {Hughes}, {Humensky}, {Kaaret}, {Karlsson}, {Kertzman},
  {Khassen}, {Kieda}, {Krawczynski}, {Krennrich}, {Lang}, {Lee}, {Madhavan},
  {Maier}, {Majumdar}, {McArthur}, {McCann}, {Millis}, {Moriarty}, {Mukherjee},
  {Nelson}, {O'Faol{\'a}in de Bhr{\'o}ithe}, {Ong}, {Orr}, {Otte}, {Pandel},
  {Park}, {Perkins}, {Pohl}, {Popkow}, {Prokoph}, {Quinn}, {Ragan}, {Reyes},
  {Reynolds}, {Roache}, {Rose}, {Ruppel}, {Saxon}, {Schroedter}, {Sembroski},
  {{\c{S}}ent{\"u}rk}, {Skole}, {Telezhinsky}, {Te{\v{s}}i{\'c}}, {Theiling},
  {Thibadeau}, {Tsurusaki}, {Tyler}, {Varlotta}, {Vassiliev}, {Vincent},
  {Wakely}, {Ward}, {Weekes}, {Weinstein}, {Weisgarber}, {Welsing}, {Williams},
  \& {Zitzer}}]{2013ApJ...770...93A}
{Aliu}, E., {Archambault}, S., {Arlen}, T., {et~al.} 2013, \apj, 770, 93

\bibitem[{{Atwood} {et~al.}(2009){Atwood}, {Abdo}, {Ackermann}, {Althouse},
  {Anderson}, {Axelsson}, {Baldini}, {Ballet}, {Band}, {Barbiellini},
  {Bartelt}, {Bastieri}, {Baughman}, {Bechtol}, {B{\'e}d{\'e}r{\`e}de},
  {Bellardi}, {Bellazzini}, {Berenji}, {Bignami}, {Bisello}, {Bissaldi},
  {Blandford}, {Bloom}, {Bogart}, {Bonamente}, {Bonnell}, {Borgland},
  {Bouvier}, {Bregeon}, {Brez}, {Brigida}, {Bruel}, {Burnett}, {Busetto},
  {Caliandro}, {Cameron}, {Caraveo}, {Carius}, {Carlson}, {Casandjian},
  {Cavazzuti}, {Ceccanti}, {Cecchi}, {Charles}, {Chekhtman}, {Cheung},
  {Chiang}, {Chipaux}, {Cillis}, {Ciprini}, {Claus}, {Cohen-Tanugi},
  {Condamoor}, {Conrad}, {Corbet}, {Corucci}, {Costamante}, {Cutini}, {Davis},
  {Decotigny}, {DeKlotz}, {Dermer}, {de Angelis}, {Digel}, {do Couto e Silva},
  {Drell}, {Dubois}, {Dumora}, {Edmonds}, {Fabiani}, {Farnier}, {Favuzzi},
  {Flath}, {Fleury}, {Focke}, {Funk}, {Fusco}, {Gargano}, {Gasparrini},
  {Gehrels}, {Gentit}, {Germani}, {Giebels}, {Giglietto}, {Giommi}, {Giordano},
  {Glanzman}, {Godfrey}, {Grenier}, {Grondin}, {Grove}, {Guillemot}, {Guiriec},
  {Haller}, {Harding}, {Hart}, {Hays}, {Healey}, {Hirayama}, {Hjalmarsdotter},
  {Horn}, {Hughes}, {J{\'o}hannesson}, {Johansson}, {Johnson}, {Johnson},
  {Johnson}, {Johnson}, {Kamae}, {Katagiri}, {Kataoka}, {Kavelaars}, {Kawai},
  {Kelly}, {Kerr}, {Klamra}, {Kn{\"o}dlseder}, {Kocian}, {Komin}, {Kuehn},
  {Kuss}, {Landriu}, {Latronico}, {Lee}, {Lee}, {Lemoine-Goumard}, {Lionetto},
  {Longo}, {Loparco}, {Lott}, {Lovellette}, {Lubrano}, {Madejski}, {Makeev},
  {Marangelli}, {Massai}, {Mazziotta}, {McEnery}, {Menon}, {Meurer},
  {Michelson}, {Minuti}, {Mirizzi}, {Mitthumsiri}, {Mizuno}, {Moiseev},
  {Monte}, {Monzani}, {Moretti}, {Morselli}, {Moskalenko}, {Murgia},
  {Nakamori}, {Nishino}, {Nolan}, {Norris}, {Nuss}, {Ohno}, {Ohsugi}, {Omodei},
  {Orlando}, {Ormes}, {Paccagnella}, {Paneque}, {Panetta}, {Parent}, {Pearce},
  {Pepe}, {Perazzo}, {Pesce-Rollins}, {Picozza}, {Pieri}, {Pinchera}, {Piron},
  {Porter}, {Poupard}, {Rain{\`o}}, {Rando}, {Rapposelli}, {Razzano}, {Reimer},
  {Reimer}, {Reposeur}, {Reyes}, {Ritz}, {Rochester}, {Rodriguez}, {Romani},
  {Roth}, {Russell}, {Ryde}, {Sabatini}, {Sadrozinski}, {Sanchez}, {Sander},
  {Sapozhnikov}, {Parkinson}, {Scargle}, {Schalk}, {Scolieri}, {Sgr{\`o}},
  {Share}, {Shaw}, {Shimokawabe}, {Shrader}, {Sierpowska-Bartosik}, {Siskind},
  {Smith}, {Smith}, {Spandre}, {Spinelli}, {Starck}, {Stephens}, {Strickman},
  {Strong}, {Suson}, {Tajima}, {Takahashi}, {Takahashi}, {Tanaka}, {Tenze},
  {Tether}, {Thayer}, {Thayer}, {Thompson}, {Tibaldo}, {Tibolla}, {Torres},
  {Tosti}, {Tramacere}, {Turri}, {Usher}, {Vilchez}, {Vitale}, {Wang},
  {Watters}, {Winer}, {Wood}, {Ylinen}, \& {Ziegler}}]{Atwood2009}
{Atwood}, W.~B., {Abdo}, A.~A., {Ackermann}, M., {et~al.} 2009, \apj, 697, 1071

\bibitem[{{Ballet} {et~al.}(2023){Ballet}, {Bruel}, {Burnett}, {Lott}, \& {The
  Fermi-LAT collaboration}}]{2023arXiv230712546B}
{Ballet}, J., {Bruel}, P., {Burnett}, T.~H., {Lott}, B., \& {The Fermi-LAT
  collaboration}. 2023, arXiv e-prints, arXiv:2307.12546

\bibitem[{{Bao} {et~al.}(2024){Bao}, {Giacinti}, {Liu}, {Zhang}, \&
  {Chen}}]{2024arXiv240702478B}
{Bao}, Y., {Giacinti}, G., {Liu}, R.-Y., {Zhang}, H.-M., \& {Chen}, Y. 2024,
  arXiv e-prints, arXiv:2407.02478

\bibitem[{{Bell}(2015)}]{2015MNRAS.447.2224B}
{Bell}, A.~R. 2015, \mnras, 447, 2224

\bibitem[{{Blasi}(2013)}]{2013A&ARv..21...70B}
{Blasi}, P. 2013, \aapr, 21, 70

\bibitem[{{Bolatto} {et~al.}(2013){Bolatto}, {Wolfire}, \&
  {Leroy}}]{bolatto2013}
{Bolatto}, A.~D., {Wolfire}, M., \& {Leroy}, A.~K. 2013, \araa, 51, 207

\bibitem[{{Dame} {et~al.}(2001){Dame}, {Hartmann}, \&
  {Thaddeus}}]{2001ApJ...547..792D}
{Dame}, T.~M., {Hartmann}, D., \& {Thaddeus}, P. 2001, \apj, 547, 792

\bibitem[{{Di Mauro} {et~al.}(2020){Di Mauro}, {Manconi}, \&
  {Donato}}]{2020PhRvD.101j3035D}
{Di Mauro}, M., {Manconi}, S., \& {Donato}, F. 2020, \prd, 101, 103035

\bibitem[{{Edwards} {et~al.}(2006){Edwards}, {Hobbs}, \&
  {Manchester}}]{2006MNRAS.372.1549E}
{Edwards}, R.~T., {Hobbs}, G.~B., \& {Manchester}, R.~N. 2006, \mnras, 372,
  1549

\bibitem[{{Fraija} \& {Araya}(2016)}]{2016ApJ...826...31F}
{Fraija}, N. \& {Araya}, M. 2016, \apj, 826, 31

\bibitem[{{Giacinti} {et~al.}(2012){Giacinti}, {Kachelrie{\ss}}, \&
  {Semikoz}}]{2012PhRvL.108z1101G}
{Giacinti}, G., {Kachelrie{\ss}}, M., \& {Semikoz}, D.~V. 2012, \prl, 108,
  261101

\bibitem[{{Giacinti} {et~al.}(2013){Giacinti}, {Kachelrie{\ss}}, \&
  {Semikoz}}]{2013PhRvD..88b3010G}
{Giacinti}, G., {Kachelrie{\ss}}, M., \& {Semikoz}, D.~V. 2013, \prd, 88,
  023010

\bibitem[{{Giacinti} {et~al.}(2020){Giacinti}, {Mitchell}, {L{\'o}pez-Coto},
  {Joshi}, {Parsons}, \& {Hinton}}]{2020A&A...636A.113G}
{Giacinti}, G., {Mitchell}, A.~M.~W., {L{\'o}pez-Coto}, R., {et~al.} 2020,
  \aap, 636, A113

\bibitem[{{Giuliani} {et~al.}(2011){Giuliani}, {Cardillo}, {Tavani}, {Fukui},
  {Yoshiike}, {Torii}, {Dubner}, {Castelletti}, {Barbiellini}, {Bulgarelli},
  {Caraveo}, {Costa}, {Cattaneo}, {Chen}, {Contessi}, {Del Monte},
  {Donnarumma}, {Evangelista}, {Feroci}, {Gianotti}, {Lazzarotto}, {Lucarelli},
  {Longo}, {Marisaldi}, {Mereghetti}, {Pacciani}, {Pellizzoni}, {Piano},
  {Picozza}, {Pittori}, {Pucella}, {Rapisarda}, {Rappoldi}, {Sabatini},
  {Soffitta}, {Striani}, {Trifoglio}, {Trois}, {Vercellone}, {Verrecchia},
  {Vittorini}, {Colafrancesco}, {Giommi}, \& {Bignami}}]{2011ApJ...742L..30G}
{Giuliani}, A., {Cardillo}, M., {Tavani}, M., {et~al.} 2011, \apjl, 742, L30

\bibitem[{{Gottschalk} {et~al.}(2012){Gottschalk}, {Kothes}, {Matthews},
  {Landecker}, \& {Dent}}]{2012A&A...541A..79G}
{Gottschalk}, M., {Kothes}, R., {Matthews}, H.~E., {Landecker}, T.~L., \&
  {Dent}, W.~R.~F. 2012, \aap, 541, A79

\bibitem[{{H.~E.~S.~S. Collaboration} {et~al.}(2018){H.~E.~S.~S.
  Collaboration}, {Abdalla}, {Abramowski}, {Aharonian}, {Ait Benkhali},
  {Akhperjanian}, {Andersson}, {Ang{\"u}ner}, {Arrieta}, {Aubert}, {Backes},
  {Balzer}, {Barnard}, {Becherini}, {Becker Tjus}, {Berge}, {Bernhard},
  {Bernl{\"o}hr}, {Blackwell}, {B{\"o}ttcher}, {Boisson}, {Bolmont}, {Bordas},
  {Bregeon}, {Brun}, {Brun}, {Bryan}, {Bulik}, {Capasso}, {Carr}, {Carrigan},
  {Casanova}, {Cerruti}, {Chakraborty}, {Chalme-Calvet}, {Chaves}, {Chen},
  {Chevalier}, {Chr{\'e}tien}, {Colafrancesco}, {Cologna}, {Condon}, {Conrad},
  {Couturier}, {Cui}, {Davids}, {Degrange}, {Deil}, {Devin}, {deWilt},
  {Dirson}, {Djannati-Ata{\"\i}}, {Domainko}, {Donath}, {Drury}, {Dubus},
  {Dutson}, {Dyks}, {Edwards}, {Egberts}, {Eger}, {Ernenwein}, {Eschbach},
  {Farnier}, {Fegan}, {Fernandes}, {Fiasson}, {Fontaine}, {F{\"o}rster},
  {Funk}, {F{\"u}{\ss}ling}, {Gabici}, {Gajdus}, {Gallant}, {Garrigoux},
  {Giavitto}, {Giebels}, {Glicenstein}, {Gottschall}, {Goyal}, {Grondin},
  {Hadasch}, {Hahn}, {Haupt}, {Hawkes}, {Heinzelmann}, {Henri}, {Hermann},
  {Hervet}, {Hillert}, {Hinton}, {Hofmann}, {Hoischen}, {Holler}, {Horns},
  {Ivascenko}, {Jacholkowska}, {Jamrozy}, {Janiak}, {Jankowsky}, {Jankowsky},
  {Jingo}, {Jogler}, {Jouvin}, {Jung-Richardt}, {Kastendieck},
  {Katarzy{\'n}ski}, {Katz}, {Kerszberg}, {Kh{\'e}lifi}, {Kieffer}, {King},
  {Klepser}, {Klochkov}, {Klu{\'z}niak}, {Kolitzus}, {Komin}, {Kosack},
  {Krakau}, {Kraus}, {Krayzel}, {Kr{\"u}ger}, {Laffon}, {Lamanna}, {Lau},
  {Lees}, {Lefaucheur}, {Lefranc}, {Lemi{\`e}re}, {Lemoine-Goumard}, {Lenain},
  {Leser}, {Lohse}, {Lorentz}, {Liu}, {L{\'o}pez-Coto}, {Lypova}, {Marandon},
  {Marcowith}, {Mariaud}, {Marx}, {Maurin}, {Maxted}, {Mayer}, {Meintjes},
  {Meyer}, {Mitchell}, {Moderski}, {Mohamed}, {Mohrmann}, {Mor{\r{a}}},
  {Moulin}, {Murach}, {de Naurois}, {Niederwanger}, {Niemiec}, {Oakes},
  {O'Brien}, {Odaka}, {{\"O}ttl}, {Ohm}, {de O{\~n}a Wilhelmi}, {Ostrowski},
  {Oya}, {Padovani}, {Panter}, {Parsons}, {Paz Arribas}, {Pekeur}, {Pelletier},
  {Perennes}, {Petrucci}, {Peyaud}, {Pita}, {Poon}, {Prokhorov}, {Prokoph},
  {P{\"u}hlhofer}, {Punch}, {Quirrenbach}, {Raab}, {Reimer}, {Reimer},
  {Renaud}, {de los Reyes}, {Rieger}, {Romoli}, {Rosier-Lees}, {Rowell},
  {Rudak}, {Rulten}, {Sahakian}, {Salek}, {Sanchez}, {Santangelo}, {Sasaki},
  {Schlickeiser}, {Sch{\"u}ssler}, {Schulz}, {Schwanke}, {Schwemmer},
  {Settimo}, {Seyffert}, {Shafi}, {Shilon}, {Simoni}, {Sol}, {Spanier},
  {Spengler}, {Spies}, {Stawarz}, {Steenkamp}, {Stegmann}, {Stinzing}, {Stycz},
  {Sushch}, {Tavernet}, {Tavernier}, {Taylor}, {Terrier}, {Tibaldo}, {Tiziani},
  {Tluczykont}, {Trichard}, {Tuffs}, {Uchiyama}, {Valerius}, {van der Walt},
  {van Eldik}, {van Soelen}, {Vasileiadis}, {Veh}, {Venter}, {Viana},
  {Vincent}, {Vink}, {Voisin}, {V{\"o}lk}, {Vuillaume}, {Wadiasingh}, {Wagner},
  {Wagner}, {Wagner}, {White}, {Wierzcholska}, {Willmann}, {W{\"o}rnlein},
  {Wouters}, {Yang}, {Zabalza}, {Zaborov}, {Zacharias}, {Zdziarski}, {Zech},
  {Zefi}, {Ziegler}, \& {{\.Z}ywucka}}]{2018A&A...612A...2H}
{H.~E.~S.~S. Collaboration}, {Abdalla}, H., {Abramowski}, A., {et~al.} 2018,
  \aap, 612, A2

\bibitem[{{H.~E.~S.~S. Collaboration} {et~al.}(2023){H.~E.~S.~S.
  Collaboration}, {Aharonian}, {Ait Benkhali}, {Aschersleben}, {Ashkar},
  {Backes}, {Barbosa Martins}, {Batzofin}, {Becherini}, {Berge},
  {Bernl{\"o}hr}, {Bi}, {B{\"o}ttcher}, {Boisson}, {Bolmont}, {Borowska},
  {Bouyahiaoui}, {Bradascio}, {Brose}, {Brun}, {Bruno}, {Bulik},
  {Burger-Scheidlin}, {Cangemi}, {Caroff}, {Casanova}, {Celic}, {Cerruti},
  {Chambery}, {Chand}, {Chandra}, {Chen}, {Chibueze}, {Chibueze}, {Cotter},
  {Mbarubucyeye}, {Devin}, {Djannati-Ata{\"\i}}, {Dmytriiev}, {Egberts},
  {Einecke}, {Ernenwein}, {Feijen}, {Fichet de Clairfontaine}, {Filipovic},
  {Fontaine}, {F{\"u}{\ss}ling}, {Funk}, {Gabici}, {Gallant}, {Ghafourizadeh},
  {Giavitto}, {Giunti}, {Glawion}, {Glicenstein}, {Goswami}, {Grolleron},
  {Grondin}, {Haerer}, {Haupt}, {Hermann}, {Hinton}, {Hofmann}, {Holch},
  {Holler}, {Horns}, {Huang}, {Jamrozy}, {Jankowsky}, {Joshi}, {Jung-Richardt},
  {Kasai}, {Katarzy{\'n}ski}, {Kh{\'e}lifi}, {Klu{\'z}niak}, {Komin}, {Kosack},
  {Kostunin}, {Lang}, {Le Stum}, {Leitl}, {Lemi{\`e}re}, {Lemoine-Goumard},
  {Lenain}, {Leuschner}, {Lohse}, {Luashvili}, {Lypova}, {Mackey}, {Malyshev},
  {Marandon}, {Marchegiani}, {Marcowith}, {Marinos}, {Mart{\'\i}-Devesa},
  {Marx}, {Maurin}, {Meintjes}, {Meyer}, {Mitchell}, {Moderski}, {Mohrmann},
  {Montanari}, {Moulin}, {Muller}, {Nakashima}, {de Naurois}, {Niemiec},
  {Noel}, {O'Brien}, {Ohm}, {Olivera-Nieto}, {de Ona Wilhelmi}, {Ostrowski},
  {Panny}, {Panter}, {Parsons}, {Peron}, {Prokhorov}, {P{\"u}hlhofer},
  {Quirrenbach}, {Reimer}, {Reimer}, {Renaud}, {Reville}, {Rieger}, {Rowell},
  {Rudak}, {Ricarte}, {Ruiz-Velasco}, {Sahakian}, {Salzmann}, {Santangelo},
  {Sasaki}, {Sch{\"u}ssler}, {Schutte}, {Schwanke}, {Shapopi}, {Sinha}, {Sol},
  {Specovius}, {Spencer}, {Stawarz}, {Steinmassl}, {Sushch}, {Suzuki},
  {Takahashi}, {Tanaka}, {Tavernier}, {Taylor}, {Terrier}, {Thorpe-Morgan},
  {Tsirou}, {Tsuji}, {Vecchi}, {Venter}, {Vink}, {Wagner}, {White},
  {Wierzcholska}, {Wong}, {Zacharias}, {Zargaryan}, {Zdziarski}, {Zech},
  {Zouari}, \& {{\.Z}ywucka}}]{2023A&A...673A.148H}
{H.~E.~S.~S. Collaboration}, {Aharonian}, F., {Ait Benkhali}, F., {et~al.}
  2023, \aap, 673, A148

\bibitem[{{Hanabata} {et~al.}(2014){Hanabata}, {Katagiri}, {Hewitt}, {Ballet},
  {Fukazawa}, {Fukui}, {Hayakawa}, {Lemoine-Goumard}, {Pedaletti}, {Strong},
  {Torres}, \& {Yamazaki}}]{hanabata2014detailed}
{Hanabata}, Y., {Katagiri}, H., {Hewitt}, J.~W., {et~al.} 2014, \apj, 786, 145

\bibitem[{{Helene}(1983)}]{helene1983}
{Helene}, O. 1983, Nuclear Instruments and Methods in Physics Research, 212,
  319

\bibitem[{{Hillas}(2005)}]{Hillas2005}
{Hillas}, A.~M. 2005, Journal of Physics G Nuclear Physics, 31, R95

\bibitem[{{Hobbs} {et~al.}(2006){Hobbs}, {Edwards}, \&
  {Manchester}}]{2006MNRAS.369..655H}
{Hobbs}, G.~B., {Edwards}, R.~T., \& {Manchester}, R.~N. 2006, \mnras, 369, 655

\bibitem[{{Jogler} \& {Funk}(2016)}]{2016ApJ...816..100J}
{Jogler}, T. \& {Funk}, S. 2016, \apj, 816, 100

\bibitem[{{Kraichnan}(1965)}]{1965PhFl....8.1385K}
{Kraichnan}, R.~H. 1965, Physics of Fluids, 8, 1385

\bibitem[{{Ladouceur} \& {Pineault}(2008)}]{2008A&A...490..197L}
{Ladouceur}, Y. \& {Pineault}, S. 2008, \aap, 490, 197

\bibitem[{{Lande} {et~al.}(2012){Lande}, {Ackermann}, {Allafort}, {Ballet},
  {Bechtol}, {Burnett}, {Cohen-Tanugi}, {Drlica-Wagner}, {Funk}, {Giordano},
  {Grondin}, {Kerr}, \& {Lemoine-Goumard}}]{lande2012}
{Lande}, J., {Ackermann}, M., {Allafort}, A., {et~al.} 2012, \apj, 756, 5

\bibitem[{{Li} \& {Chen}(2010)}]{li2010gamma}
{Li}, H. \& {Chen}, Y. 2010, \mnras, 409, L35

\bibitem[{{Li} {et~al.}(2024){Li}, {Liu}, \& {Giacinti}}]{2024A&A...689A.257L}
{Li}, Y., {Liu}, S., \& {Giacinti}, G. 2024, \aap, 689, A257

\bibitem[{{Li} {et~al.}(2023{\natexlab{a}}){Li}, {Liu}, \&
  {He}}]{2023ApJ...953..100L}
{Li}, Y., {Liu}, S., \& {He}, Y. 2023{\natexlab{a}}, \apj, 953, 100

\bibitem[{{Li} {et~al.}(2023{\natexlab{b}}){Li}, {Xin}, {Liu}, \&
  {He}}]{2023ApJ...945...21L}
{Li}, Y., {Xin}, Y., {Liu}, S., \& {He}, Y. 2023{\natexlab{b}}, \apj, 945, 21

\bibitem[{{Liu} {et~al.}(2022){Liu}, {Zeng}, {Xin}, \&
  {Zhang}}]{2022RvMPP...6...19L}
{Liu}, S., {Zeng}, H., {Xin}, Y., \& {Zhang}, Y. 2022, Reviews of Modern Plasma
  Physics, 6, 19

\bibitem[{{Liu} {et~al.}(2020){Liu}, {Zeng}, {Xin}, \&
  {Zhu}}]{2020ApJ...897L..34L}
{Liu}, S., {Zeng}, H., {Xin}, Y., \& {Zhu}, H. 2020, \apjl, 897, L34

\bibitem[{{L{\'o}pez-Coto} {et~al.}(2022){L{\'o}pez-Coto}, {de O{\~n}a
  Wilhelmi}, {Aharonian}, {Amato}, \& {Hinton}}]{2022NatAs...6..199L}
{L{\'o}pez-Coto}, R., {de O{\~n}a Wilhelmi}, E., {Aharonian}, F., {Amato}, E.,
  \& {Hinton}, J. 2022, Nature Astronomy, 6, 199

\bibitem[{{L{\'o}pez-Coto} \& {Giacinti}(2018)}]{2018MNRAS.479.4526L}
{L{\'o}pez-Coto}, R. \& {Giacinti}, G. 2018, \mnras, 479, 4526

\bibitem[{{MAGIC Collaboration} {et~al.}(2023){MAGIC Collaboration}, {Acciari},
  {Ansoldi}, {Antonelli}, {Arbet Engels}, {Baack}, {Babi{\'c}}, {Banerjee},
  {Barres de Almeida}, {Barrio}, {Becerra Gonz{\'a}lez}, {Bednarek},
  {Bellizzi}, {Bernardini}, {Berti}, {Besenrieder}, {Bhattacharyya},
  {Bigongiari}, {Biland}, {Blanch}, {Bonnoli}, {Bo{\v{s}}njak}, {Busetto},
  {Carosi}, {Ceribella}, {Cerruti}, {Chai}, {Chilingarian}, {Cikota}, {Colak},
  {Colin}, {Colombo}, {Contreras}, {Cortina}, {Covino}, {D'Elia}, {da Vela},
  {Dazzi}, {de Angelis}, {de Lotto}, {Delfino}, {Delgado}, {Depaoli}, {di
  Pierro}, {di Venere}, {Do Souto Espi{\~n}eira}, {Dominis Prester}, {Donini},
  {Dorner}, {Doro}, {Elsaesser}, {Fallah Ramazani}, {Fattorini}, {Ferrara},
  {Foffano}, {Fonseca}, {Font}, {Fruck}, {Fukami}, {Garc{\'\i}a L{\'o}pez},
  {Garczarczyk}, {Gasparyan}, {Gaug}, {Giglietto}, {Giordano}, {Gliwny},
  {Godinovi{\'c}}, {Green}, {Hadasch}, {Hahn}, {Herrera}, {Hoang}, {Hrupec},
  {H{\"u}tten}, {Inada}, {Inoue}, {Ishio}, {Iwamura}, {Jouvin}, {Kajiwara},
  {Karjalainen}, {Kerszberg}, {Kobayashi}, {Kubo}, {Kushida}, {Lamastra},
  {Lelas}, {Leone}, {Lindfors}, {Lombardi}, {Longo}, {L{\'o}pez},
  {L{\'o}pez-Coto}, {L{\'o}pez-Oramas}, {Loporchio}, {Machado de Oliveira
  Fraga}, {Masuda}, {Maggio}, {Majumdar}, {Makariev}, {Mallamaci}, {Maneva},
  {Manganaro}, {Mannheim}, {Maraschi}, {Mariotti}, {Mart{\'\i}nez}, {Mazin},
  {Mender}, {Mi{\'c}anovi{\'c}}, {Miceli}, {Miener}, {Minev}, {Miranda},
  {Mirzoyan}, {Molina}, {Moralejo}, {Morcuende}, {Moreno}, {Moretti},
  {Munar-Adrover}, {Neustroev}, {Nigro}, {Nilsson}, {Ninci}, {Nishijima},
  {Noda}, {Nogu{\'e}s}, {Nozaki}, {Ohtani}, {Oka}, {Otero-Santos},
  {Palatiello}, {Paneque}, {Paoletti}, {Paredes}, {Pavleti{\'c}}, {Pe{\~n}il},
  {Peresano}, {Persic}, {Prada Moroni}, {Prandini}, {Puljak}, {Rhode},
  {Rib{\'o}}, {Rico}, {Righi}, {Rugliancich}, {Saha}, {Sahakyan}, {Saito},
  {Sakurai}, {Satalecka}, {Schleicher}, {Schmidt}, {Schweizer}, {Sitarek},
  {{\v{S}}nidari{\'c}}, {Sobczynska}, {Spolon}, {Stamerra}, {Strom}, {Strzys},
  {Suda}, {Suri{\'c}}, {Takahashi}, {Tavecchio}, {Temnikov}, {Terzi{\'c}},
  {Teshima}, {Torres-Alb{\`a}}, {Tosti}, {van Scherpenberg}, {Vanzo}, {Vazquez
  Acosta}, {Ventura}, {Verguilov}, {Vigorito}, {Vitale}, {Vovk}, {Will},
  {Zari{\'c}}, {Celli}, \& {Morlino}}]{2023A&A...670A...8M}
{MAGIC Collaboration}, {Acciari}, V.~A., {Ansoldi}, S., {et~al.} 2023, \aap,
  670, A8

\bibitem[{{Mares} {et~al.}(2021){Mares}, {Lemoine-Goumard}, {Acero}, {Clark},
  {Devin}, {Gabici}, {Gelfand}, {Green}, \& {Grondin}}]{2021ApJ...912..158M}
{Mares}, A., {Lemoine-Goumard}, M., {Acero}, F., {et~al.} 2021, \apj, 912, 158

\bibitem[{{Mattox} {et~al.}(1996){Mattox}, {Bertsch}, {Chiang}, {Dingus},
  {Digel}, {Esposito}, {Fierro}, {Hartman}, {Hunter}, {Kanbach}, {Kniffen},
  {Lin}, {Macomb}, {Mayer-Hasselwander}, {Michelson}, {von Montigny},
  {Mukherjee}, {Nolan}, {Ramanamurthy}, {Schneid}, {Sreekumar}, {Thompson}, \&
  {Willis}}]{mattox1996likelihood}
{Mattox}, J.~R., {Bertsch}, D.~L., {Chiang}, J., {et~al.} 1996, \apj, 461, 396

\bibitem[{{Peron} {et~al.}(2020){Peron}, {Aharonian}, {Casanova}, {Zanin}, \&
  {Romoli}}]{peron2020gamma}
{Peron}, G., {Aharonian}, F., {Casanova}, S., {Zanin}, R., \& {Romoli}, C.
  2020, \apjl, 896, L23

\bibitem[{{Principe} {et~al.}(2020){Principe}, {Mitchell}, {Caroff}, {Hinton},
  {Parsons}, \& {Funk}}]{2020A&A...640A..76P}
{Principe}, G., {Mitchell}, A.~M.~W., {Caroff}, S., {et~al.} 2020, \aap, 640,
  A76

\bibitem[{{Razzano} {et~al.}(2023){Razzano}, {Fiori}, {Saz Parkinson},
  {Mignani}, {De Luca}, {Harding}, {Kerr}, {Marelli}, \&
  {Testa}}]{2023A&A...676A..91R}
{Razzano}, M., {Fiori}, A., {Saz Parkinson}, P.~M., {et~al.} 2023, \aap, 676,
  A91

\bibitem[{{Reid} {et~al.}(2014){Reid}, {Menten}, {Brunthaler}, {Zheng}, {Dame},
  {Xu}, {Wu}, {Zhang}, {Sanna}, {Sato}, {Hachisuka}, {Choi}, {Immer},
  {Moscadelli}, {Rygl}, \& {Bartkiewicz}}]{2014ApJ...783..130R}
{Reid}, M.~J., {Menten}, K.~M., {Brunthaler}, A., {et~al.} 2014, \apj, 783, 130

\bibitem[{{Reid} {et~al.}(2009){Reid}, {Menten}, {Zheng}, {Brunthaler},
  {Moscadelli}, {Xu}, {Zhang}, {Sato}, {Honma}, {Hirota}, {Hachisuka}, {Choi},
  {Moellenbrock}, \& {Bartkiewicz}}]{2009ApJ...700..137R}
{Reid}, M.~J., {Menten}, K.~M., {Zheng}, X.~W., {et~al.} 2009, \apj, 700, 137

\bibitem[{{Sano} {et~al.}(2013){Sano}, {Tanaka}, {Torii}, {Fukuda}, {Yoshiike},
  {Sato}, {Horachi}, {Kuwahara}, {Hayakawa}, {Matsumoto}, {Inoue}, {Yamazaki},
  {Inutsuka}, {Kawamura}, {Tachihara}, {Yamamoto}, {Okuda}, {Mizuno}, {Onishi},
  {Mizuno}, \& {Fukui}}]{2013ApJ...778...59S}
{Sano}, H., {Tanaka}, T., {Torii}, K., {et~al.} 2013, \apj, 778, 59

\bibitem[{{Schneider} {et~al.}(2006){Schneider}, {Bontemps}, {Simon}, {Jakob},
  {Motte}, {Miller}, {Kramer}, \& {Stutzki}}]{2006A&A...458..855S}
{Schneider}, N., {Bontemps}, S., {Simon}, R., {et~al.} 2006, \aap, 458, 855

\bibitem[{{Schneider} {et~al.}(2007){Schneider}, {Simon}, {Bontemps},
  {Comer{\'o}n}, \& {Motte}}]{2007A&A...474..873S}
{Schneider}, N., {Simon}, R., {Bontemps}, S., {Comer{\'o}n}, F., \& {Motte}, F.
  2007, \aap, 474, 873

\bibitem[{{Smith} {et~al.}(2023){Smith}, {Abdollahi}, {Ajello}, {Bailes},
  {Baldini}, {Ballet}, {Baring}, {Bassa}, {Gonzalez}, {Bellazzini}, {Berretta},
  {Bhattacharyya}, {Bissaldi}, {Bonino}, {Bottacini}, {Bregeon}, {Bruel},
  {Burgay}, {Burnett}, {Cameron}, {Camilo}, {Caputo}, {Caraveo}, {Cavazzuti},
  {Chiaro}, {Ciprini}, {Clark}, {Cognard}, {Corongiu}, {Orestano},
  {Crnogorcevic}, {Cuoco}, {Cutini}, {D'Ammando}, {de Angelis}, {DeCesar}, {De
  Gaetano}, {de Menezes}, {Deneva}, {de Palma}, {Di Lalla}, {Dirirsa}, {Di
  Venere}, {Dom{\'\i}nguez}, {Dumora}, {Fegan}, {Ferrara}, {Fiori},
  {Fleischhack}, {Flynn}, {Franckowiak}, {Freire}, {Fukazawa}, {Fusco},
  {Galanti}, {Gammaldi}, {Gargano}, {Gasparrini}, {Giacchino}, {Giglietto},
  {Giordano}, {Giroletti}, {Green}, {Grenier}, {Guillemot}, {Guiriec},
  {Gustafsson}, {Harding}, {Hays}, {Hewitt}, {Horan}, {Hou}, {Jankowski},
  {Johnson}, {Johnson}, {Johnston}, {Kataoka}, {Keith}, {Kerr}, {Kramer},
  {Kuss}, {Latronico}, {Lee}, {Li}, {Li}, {Limyansky}, {Longo}, {Loparco},
  {Lorusso}, {Lovellette}, {Lower}, {Lubrano}, {Lyne}, {Maan}, {Maldera},
  {Manchester}, {Manfreda}, {Marelli}, {Mart{\'\i}-Devesa}, {Mazziotta},
  {McEnery}, {Mereu}, {Michelson}, {Mickaliger}, {Mitthumsiri}, {Mizuno},
  {Moiseev}, {Monzani}, {Morselli}, {Negro}, {Nemmen}, {Nieder}, {Nuss},
  {Omodei}, {Orienti}, {Orlando}, {Ormes}, {Palatiello}, {Paneque},
  {Panzarini}, {Parthasarathy}, {Persic}, {Pesce-Rollins}, {Pillera}, {Poon},
  {Porter}, {Possenti}, {Principe}, {Rain{\`o}}, {Rando}, {Ransom}, {Ray},
  {Razzano}, {Razzaque}, {Reimer}, {Reimer}, {Renault-Tinacci}, {Romani},
  {S{\'a}nchez-Conde}, {Parkinson}, {Scotton}, {Serini}, {Sgr{\`o}}, {Shannon},
  {Sharma}, {Shen}, {Siskind}, {Spandre}, {Spinelli}, {Stappers}, {Stephens},
  {Suson}, {Tabassum}, {Tajima}, {Tak}, {Theureau}, {Thompson}, {Tibolla},
  {Torres}, {Valverde}, {Venter}, {Wadiasingh}, {Wang}, {Wang}, {Wang},
  {Weltevrede}, {Wood}, {Yan}, {Zaharijas}, {Zhang}, \& {Zhu}}]{smi+23}
{Smith}, D.~A., {Abdollahi}, S., {Ajello}, M., {et~al.} 2023, \apj, 958, 191

\bibitem[{{Solin} {et~al.}(2012){Solin}, {Ukkonen}, \&
  {Haikala}}]{2012A&A...542A...3S}
{Solin}, O., {Ukkonen}, E., \& {Haikala}, L. 2012, \aap, 542, A3

\bibitem[{{Su} {et~al.}(2019){Su}, {Yang}, {Zhang}, {Gong}, {Wang}, {Zhou},
  {Wang}, {Chen}, {Sun}, {Chen}, {Xu}, \& {Jiang}}]{2019ApJS..240....9S}
{Su}, Y., {Yang}, J., {Zhang}, S., {et~al.} 2019, \apjs, 240, 9

\bibitem[{{Tanaka} {et~al.}(2020){Tanaka}, {Uchida}, {Sano}, \&
  {Tsuru}}]{2020ApJ...900L...5T}
{Tanaka}, T., {Uchida}, H., {Sano}, H., \& {Tsuru}, T.~G. 2020, \apjl, 900, L5

\bibitem[{{Uchiyama} {et~al.}(2012){Uchiyama}, {Funk}, {Katagiri}, {Katsuta},
  {Lemoine-Goumard}, {Tajima}, {Tanaka}, \& {Torres}}]{uchiyama2012fermi}
{Uchiyama}, Y., {Funk}, S., {Katagiri}, H., {et~al.} 2012, \apjl, 749, L35

\bibitem[{{Wang} {et~al.}(2023){Wang}, {Takata}, {Lin}, \&
  {Tam}}]{2023arXiv230703661W}
{Wang}, H.~H., {Takata}, J., {Lin}, L.~C.~C., \& {Tam}, P. H.~T. 2023, arXiv
  e-prints, arXiv:2307.03661

\bibitem[{{Wood} {et~al.}(2017){Wood}, {Caputo}, {Charles}, {Di Mauro},
  {Magill}, {Perkins}, \& {Fermi-LAT Collaboration}}]{2017ICRC...35..824W}
{Wood}, M., {Caputo}, R., {Charles}, E., {et~al.} 2017, in International Cosmic
  Ray Conference, Vol. 301, 35th International Cosmic Ray Conference
  (ICRC2017), 824

\bibitem[{{Yuan} {et~al.}(2012){Yuan}, {Liu}, \& {Bi}}]{2012ApJ...761..133Y}
{Yuan}, Q., {Liu}, S., \& {Bi}, X. 2012, \apj, 761, 133

\bibitem[{{Zabalza}(2015)}]{zabalza2015naima}
{Zabalza}, V. 2015, in International Cosmic Ray Conference, Vol.~34, 34th
  International Cosmic Ray Conference (ICRC2015), 922

\bibitem[{{Zeng} {et~al.}(2019){Zeng}, {Xin}, \& {Liu}}]{2019ApJ...874...50Z}
{Zeng}, H., {Xin}, Y., \& {Liu}, S. 2019, \apj, 874, 50

\end{thebibliography}

\end{document}